\documentclass[aps,prb,twocolumn,superscriptaddress]{revtex4-2}

\usepackage{amsmath, amssymb,  graphicx, caption}

\usepackage[colorlinks=true ,urlcolor=blue,urlbordercolor={0 1 1}]{hyperref}

\usepackage{color}
\usepackage{xcolor}
\usepackage{braket}

\usepackage[utf8]{inputenc}
\usepackage{bm}
\usepackage{verbatim}
\usepackage{graphicx}
\usepackage{amsmath}
\usepackage{color}
\usepackage{float}
\usepackage{bbm}
\usepackage{amssymb}
\usepackage{slashed}
\usepackage{wasysym,bm,bbm,dsfont}

\begin{document}
\title{Deconfined Fermi liquid to Fermi liquid transition and superconducting instability}
\author{Xiaofan Wu }
\thanks{These two authors contributed equally}
\author{Hui Yang}
\thanks{These two authors contributed equally}
\author{Ya-Hui Zhang}
\affiliation{William H. Miller III Department of Physics and Astronomy, Johns Hopkins University, Baltimore, Maryland, 21218, USA}
\begin{abstract}

Deconfined quantum critical points (DQCP)  have attracted lots of attentions in the past decades, but were mainly restricted to incompressible phases. On the other hand, various experimental puzzles call for new theory of unconventional quantum criticality between metals at a generic density. Here we explore the possibility of a deconfined transition between two symmetric Fermi liquids (FL) in a bilayer model tuned by inter-layer antiferromagnetic spin-spin coupling $J_\perp$. Across the transition the Fermi surface volume per flavor jumps by $1/2$ of the Brillouin zone (BZ), similar to the small to large Fermi surface transitions in heavy Fermion systems and maybe also in the high Tc cuprates. But in the bilayer case the small Fermi surface phase (dubbed as sFL) has neither symmetry breaking nor fractionalization, akin to the symmetric mass generation (SMG) discussed in high energy physics. We formulate a deconfined critical theory where the two Fermi liquids correspond to higgs and/or confined phases of a  $U(1)\times U(1)$ gauge theory.  We show that this deconfined FL to FL transition (DFFT) fixed point is unstable to pairing and thus a superconductor dome is expected at low temperature. At finite temperature above the pairing scale, microscopic electron is a three particle bound state of the deconfined fractional fermions in the critical theory.  We also introduce another parameter which can suppress the pairing instability, leading to a deconfined phase stable to zero temperature. Our work opens a new direction to exploring deconfined metallic criticality and new pairing mechanism from critical gauge field. The transition may be relevant to the recently found nickelate superconductor La$_3$Ni$_2$O$_7$ and future experiments in bilayer optical lattice.
\end{abstract}
\maketitle

\section{Introduction}

There have been intensive studies on possible unconventional transitions beyond the familiar Landau-Ginzburg framework based on symmetry breaking order parameters. One famous example is the deconfined quantum critical point (DQCP)\cite{senthil2004deconfined,senthil2023deconfined,PhysRevB.99.075103,PhysRevB.99.165143,PhysRevB.100.125137} between the Neel ordered and valence bond solid (VBS) phases on square lattice. DQCP was also suggested for symmetric mass generation (SMG)\cite{wang2022symmetric,tong2022comments} transition between a semimetal and an insulator\cite{you2018symmetric}. In these examples the two sides of the phase transitions are just conventional phases without any fractionalization, but the critical regime is described by fractionalized particle and emergent internal gauge field at low energy. So far the discussions of deconfined criticalities are largely restricted to insulators or semimetals at integer filling. On the other hand, experiments in the heavy fermion systems\cite{coleman2001fermi,gegenwart2008quantum,si2010heavy,stewart2001non,coleman2005quantum,lohneysen2007fermi,senthil2005quantum,kirchner2020colloquium} and in the high temperature superconductor cuprates\cite{lee2006doping,sachdev2003colloquium,phillips2022stranger,proust2019remarkable} suggest the possibility of a quantum critical point with Fermi surface volume jump between two metallic phases. Besides, the phenomenology seems to be beyond the conventional metallic criticality simply with a fluctuating symmetry breaking order parameter such as in the Hertz-Millis-Moriya theory\cite{hertz1976quantum,millis1993effect,lohneysen2007fermi}. Therefore it is important to generalize the idea of DQCP to the more sophisticated phase transition with Fermi surfaces in the two sides. Critical theories have been proposed for the unusual case where one side has a neutral Fermi surface\cite{senthil2004weak,senthil2008theory}, but the examples are essentially Mott or orbital-selective Mott transition with the volume of the critical Fermi surface fixed at the  half filling. A DQCP with both sides as conventional metallic phases with arbitrary size of Fermi surfaces is still elusive, despite of some progress for a deconfined metallic transition with an onset of antiferromagnetism\cite{zhang2020deconfined}.

In this work we turn to a different setup with a bilayer model and identify a much cleaner small to large Fermi surface transition with symmetric Fermi liquids in both sides. More specifically, we consider a bilayer Hubbard or t-J model with strong inter-layer spin-spin interaction $J_\perp$, but no inter-layer hopping $t_\perp$. Naively this seems impossible because usually $J_\perp$ is generated from the $t^2_{\perp}/U$ super-exchange process. But it is actually possible to generate a large $J_\perp$ from Hund's coupling to a rung-singlet from a different orbital as proposed by one of us for the recently found nickelate superconductor\cite{oh2023type,lu2023interlayer,qu2023bilayer,yang2023strong,Fabian2023_pairing_dome}. In this situation, the symmetry is $(U(1)_t\times U(1)_b \times SU(2) )/Z_2$ with the two U(1) corresponding to the charge conservations of the top and bottom layers respectively. Oshikawa's non-perturbative proof of the Luttinger theorem\cite{oshikawa2000topological} then shows that there are two classes of symmetric and featureless\footnote{By featureless we mean that the phase does not have fractionalization.} Fermi liquids: a conventional Fermi liquid (FL) and a second Fermi liquid (sFL)\cite{zhang2020spin,yang2023strong}. The sFL phase has Fermi surface volume smaller than the FL phase by 1/2 of the Brillouin zone (BZ) per flavor. At a fixed density per layer $n=1-x$ with small hole doping level $x$, we have a FL phase with large Fermi surface volume $A_{FS}=\frac{1-x}{2}$ per flavor at small $J_\perp$. Then in the large $J_\perp$ regime, the sFL phase  with $A_{FS}=-\frac{x}{2}$ is stabilized instead. Therefore there is a large to small Fermi surface transition tuned by $J_\perp$. The sFL phase is clearly beyond any weak coupling theory and it arises only in the strong coupling regime of the four fermion interaction $J_\perp$ akin to the symmetric mass generation\cite{you2018symmetric,lu2023fermi} discussed in the high energy physics, though in our case the charge carriers are only partially gapped.

The FL to sFL transition, if continuous, must be beyond Landau-Ginzburg framework as there is no symmetry breaking order parameter. Meanwhile both phases are conventional without fractionalization. Thus it is natural to expect a deconfined criticality similar to the Neel to VBS DQCP. We formulate such a theory in this work. In our critical theory, the two phases correspond to higgsed and/or confined phases of a U(1)$\times$ U(1) gauge theory with deconfined fractionalized particle existing only at the critical regime.  The critical theory is unstable to pairing at zero temperature, but the pairing scale can be suppressed to be arbitrarily low due to the almost balance between the repulsive and attractive interaction from the two U(1) gauge fields.  Similar features of balance from two gauge fields have been discussed previously in other contexts\cite{zhang2020pseudogap,zhang2020deconfined,zou2020deconfined,zou2020deconfined_qh}. Above the pairing energy scale, we have a large critical regime where electron is a three particle bound state of the elementary fermions in the low energy theory. One immediate implication is that the quasiparticle residue $Z$ vanishes at the critical regime and any single electron spectroscopy measurement (such as ARPES or STM) can not see any coherent quasiparticle. We also include another axis to tune $\delta U$, the difference between the intra-layer and inter-layer repulsion. We argue that the pairing instability can be suppressed by tuning $\delta U$ and there is an intermediate deconfined metal (DM) phase which is similar to the deconfined critical regime. Our work opens a door to study DQCP between compressible phases, which, unlike the previous discussions in insulators or semimetals, can arise at any electron density.

\section{One orbital bilayer Hubbard model}

We consider the following one-orbital bilayer model:

\begin{equation}
    \begin{aligned}
    H&=-t\sum_{\langle ij \rangle}\sum_{\alpha}c^{\dagger}_{i;\alpha}c_{j;\alpha} + \frac{1}{2}U_0 \sum_i \sum_{a=t,b}n^2_{i;a}\\ 
    &+ V_0\sum_i n_{i;t}n_{i;b} + J_\perp \sum_i \vec{S}_{i;t}\cdot\vec{S}_{i;b}.
    \end{aligned}
    \label{eq:bilayer_hubbard}
\end{equation}
We view layer as a pseudospin, then we have four flavors labeled as $\alpha=a,\sigma$. $a=t,b$ labels the top and bottom layer while $\sigma=\uparrow,\downarrow$ labels the spin.  $n_{i;t}$ and $n_{i;b}$ indicate the density at site $i$ for the top and bottom layers respectively. $\vec S_{i;t}$ and $\vec S_{i;b}$ are the spin operators in the two layers. If $V_0=U_0$ and $J_\perp=0$, this is the SU(4) Hubbard model. Generally $U_0$ and $V_0$ can be different and the model can be rewritten as:

\begin{equation}
    \label{eq:general_SU4_model}
    \begin{aligned}
    H&=-t\sum_{\langle ij \rangle}\sum_{\alpha}c^{\dagger}_{i;\alpha}c_{j;\alpha}+\frac{1}{2}U\sum_i n_i^2 \\
    &+\delta U \sum_i P_{i;z}^2 + J_\perp \sum_i \vec{S}_{i;t}\cdot\vec{S}_{i;b},
    \end{aligned}
\end{equation}
where $n_i = n_{i;t} + n_{i;b}$ and $P_{i;z} = n_{i;t} - n_{i;b}$. We have $U = \frac{1}{2}\left( U_0 + V_0 \right)$ and $\delta U = \frac{1}{4}\left( U_0 - V_0 \right)$. The model has a $\left( U(1)_t\times U(1)_b\times SU(2) \right)/Z_2$ symmetry with $U(1)$ charge conservation in the two layers separately because there is no $t_\perp$ hopping. We have electron density(summed over spin) per layer to be $n_t=n_b=1-x$ per site. So the filling per spin per layer is $\nu=\frac{1-x}{2}$.

\begin{figure}
    \centering
    \includegraphics[scale=0.6]{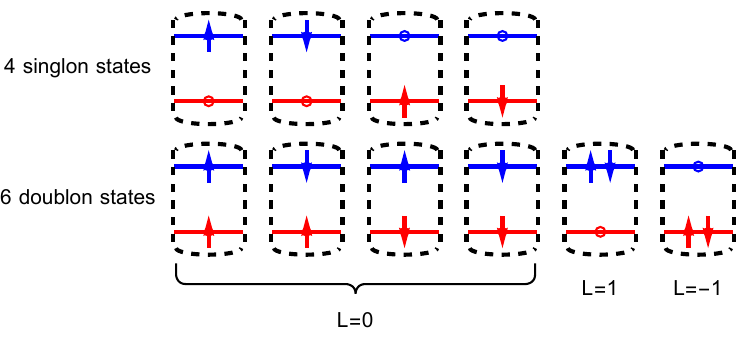}
    \caption{Restricted Hilbert space at each site in the large $U$ regime.  The dashed box represents a single site (combine the two layers). Blue lines and red lines represent top and bottom layer, respectively. The empty state is penalized by the large $U$ and there are 4 singlon states and 6 doublon states. The last 2 doublon states are further penalized by $\delta U$ and should be removed if we also take the large $\delta U$ limit.}
    \label{fig:4+6_hilbert_space}
\end{figure}

We are interested in the large $U_0$ and $V_0$ regime in this work with $U_0>V_0\gg t,J_\perp$. Note that the electron density summed over two layers and spin is $n_T = 2(1-x)$. $n_T = 1$ corresponds to the Mott insulator and we consider the regime $1<n_T < 2$. The restricted Hilbert space due to the large U is shown in Fig.~(\ref{fig:4+6_hilbert_space}) which consists of 4 singlon states (with $n_T=1$) and 6 doublon states (with $n_T=2$). The empty state is forbidden since we need to add one more doubly occupied site to create one empty site, which costs energy $U_0$ or $V_0$. The Hamiltonian in this restricted Hilbert space now becomes\cite{zhang2020spin}

\begin{equation}
    \label{eq:4+6_t_J_model}
    \begin{aligned}
        H &= -t\sum_{\langle ij \rangle}\sum_\alpha P_{4+6}c^\dagger_{i;\alpha}c_{j;\alpha} P_{4+6} \\
        &+ J_\perp \sum_i \vec{S}_{i;t}\cdot \vec{S}_{i;b} 
        +\delta U \sum_i P_{i;z}^2+...,
    \end{aligned}
\end{equation}
where $P_{4+6}$ is the projection operator into the 4-singlon-6-doublon Hilbert space shown in Fig.~\ref{fig:4+6_hilbert_space}. Note that here we include the intra-layer spin-spin coupling in the $...$ term.

If we further take the limit that $\delta U$ is also large, the last two doublon states in Fig.~(\ref{fig:4+6_hilbert_space}) are also forbidden. Then the restricted Hilbert space on each site now has 4 single occupied states and 4 double occupied states. The Hamiltonian in this restricted Hilbert space simply consists of $t$ and $J_\perp$ term:

\begin{equation}
    \label{eq:4+4_tJ_model}
    \begin{aligned}
    H &= -t\sum_{\langle ij \rangle}\sum_\alpha P_{4+4}c^\dagger_{i;\alpha} c_{j;\alpha}P_{4+4} \\
    &+ J_\perp \sum_i \vec{S}_{i;t}\cdot \vec{S}_{i;b}+...,
    \end{aligned}
\end{equation}
where $P_{4+4}$ is the projection operator into the 4-singlon-4-doublon Hilbert space. Again we include the intra-layer Heisenberg spin-spin coupling in the $...$ term.

\subsection{Relation to the bilayer nickelate}

The model in Eq.~\ref{eq:bilayer_hubbard} is unusual in the sense that there is a $J_\perp$, but no inter-layer hopping $t_\perp$. As usually spin-spin coupling is from second order super-exchange process with $J_\perp \sim \frac{t_\perp^2}{U}$, it is not clear that the model is physical.  The model has been discussed by one of us in the context of graphene moir\'e system\cite{zhang2020spin} where the valley plays the role of the layer and the $J_\perp$ term is from the phonon mediated anti-Hund's coupling. The model was also proposed for bilayer optical lattice with strong inter-layer potential difference, though in a non-equilibrium setting\cite{bohrdt2022strong}.

\begin{figure}
    \centering
    \includegraphics[scale=0.35]{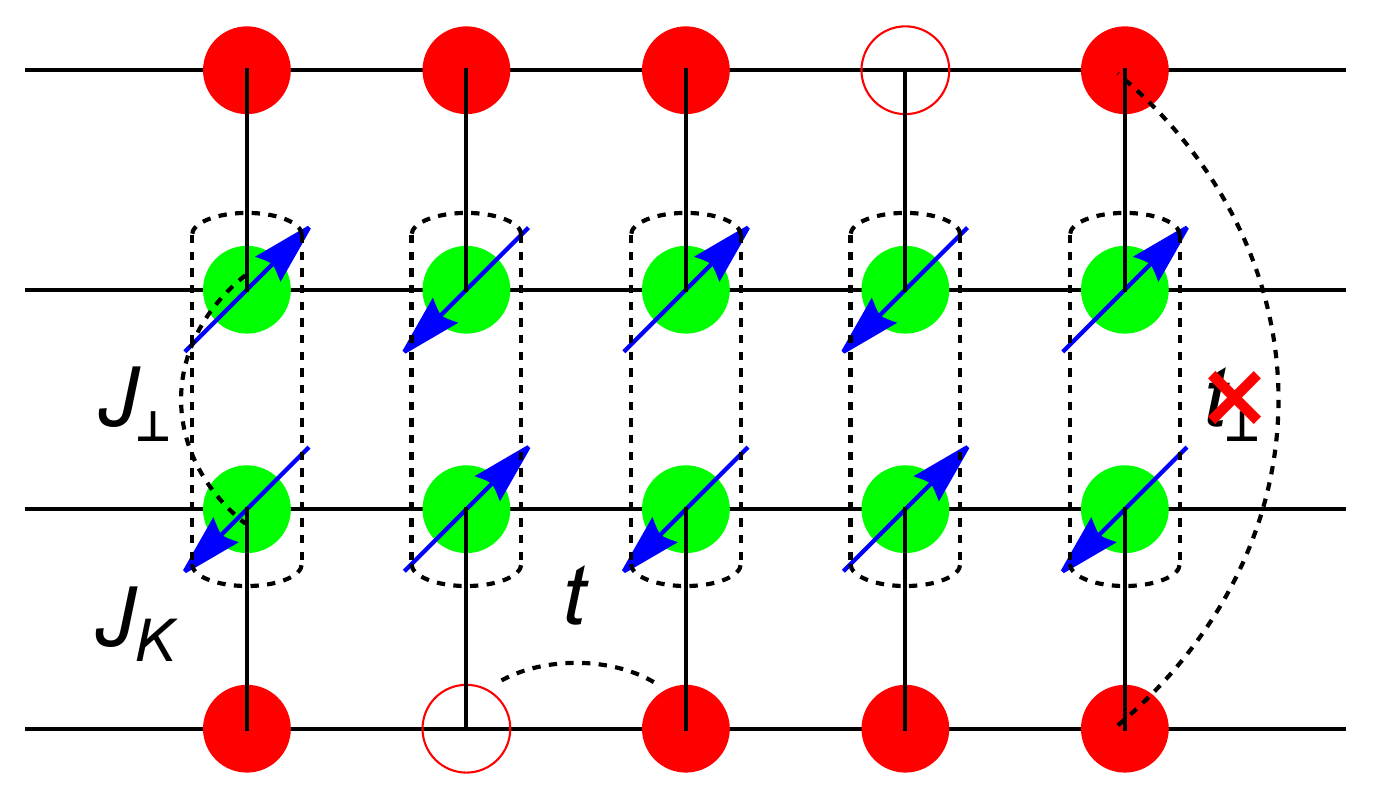}
    \caption{Illustration of the double Kondo model for the bilayer nickelate. The solid green circle with arrow, solid red circle, and the red circle correspond to the spin-$1/2$ moments, electron, and hole, respectively. $t$, $J_\perp$, $J_K$ are the hopping, inter-layer coupling and the Kondo coupling (Hund's coupling). There is no $t_\perp$ in this model. In the large $J_\perp$ limit, the spin moments form rung singlet $\frac{1}{\sqrt{2}}(\ket{\uparrow\downarrow}-\ket{\downarrow\uparrow})$.}
    \label{fig:double_kondo}
\end{figure}

More recently a more realistic realization of the model has been proposed\cite{oh2023type,lu2023interlayer,qu2023bilayer,yang2023strong,Fabian2023_pairing_dome} for the recently found nickelate superconductor La$_3$Ni$_2$O$_7$ under pressure with $n_t=n_b=1-x$ and $x \approx 0.5$.  The key is  to have additional spin moments forming bilayer rung singlet. Then the itinerant electron couples to the these spin moments through Hund's coupling or super-exchange coupling $J_K$, which shares the strong $J_\perp$ of the spin moments to the itinerant electron.  In the end the itinerant electron feels a strong $J_\perp$, but without $t_\perp$ as shown in Fig.~\ref{fig:double_kondo}.

In this situation one expect that $V_0\ll U_0$ and $U_0 \gg t$.  In realistic system $V_0$ may not be too large, therefore one should also keep the empty state for each rung in the low energy Hilbert space, as is done in Ref.~\onlinecite{yang2023strong} by two of us. From our previous analysis at finite $V_0$, there are two different normal states: the conventional Fermi liquid and the second Fermi liquid (sFL) with Fermi surface volume smaller by 1/2 of the BZ per flavor. The sFL still satisfies Oshikawa's non-perturbative proof of the Luttinger theorem, despite that it is beyond any weak coupling theory and is intrinsically strongly correlated. At low temperature, both the FL and sFL are unstable to superconductivity due to an attractive interaction mediated by an on-site virtual Cooper pair. In this work we are interested in the transition between the FL and the sFL phase, so we will make $V_0 \rightarrow +\infty$ to suppress the pairing instability discussed in Ref.~\onlinecite{yang2023strong} completely.\\

\begin{figure}[ht]
    \centering        \includegraphics[width=0.5\linewidth]{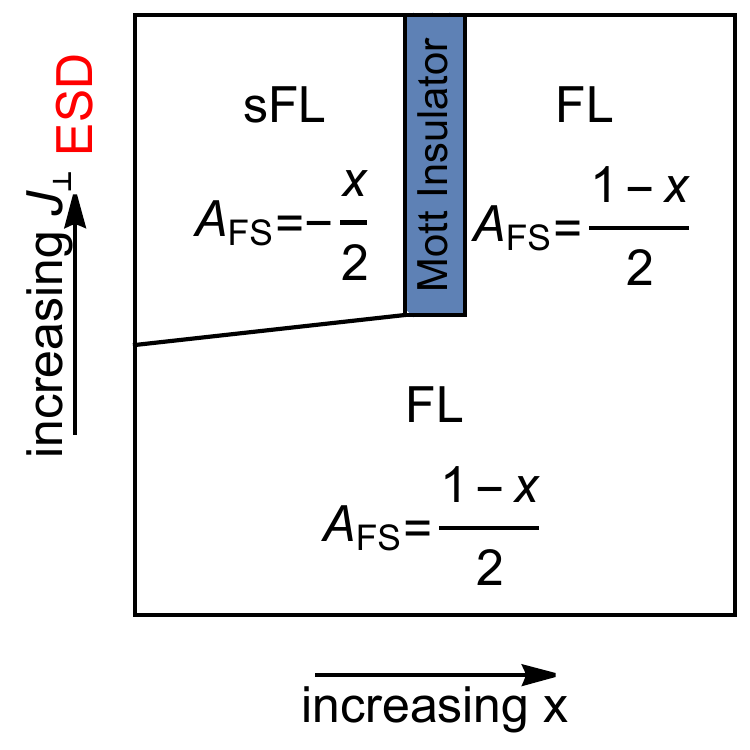}
    \caption{Illustrated phases of the bilayer model with doping and $J_\perp$, for small $J_\perp$, it is always a FL, while for large $J_\perp$, we can reduce the bilayer model to the ESD model, with a sFL in small $x$ and a FL in large $x$.
    }
    \label{fig:bilayer_phase}
\end{figure}

In Fig.~\ref{fig:bilayer_phase}, we show the illustrated phase diagram of model Eq.~\ref{eq:4+6_t_J_model} and Eq.~\ref{eq:4+4_tJ_model}. In the small $J_\perp$ limit, as we increase the doping, the system is always in the FL phase with Fermi surface volume $A_{FS}=\frac{1-x}{2}$ per flavor. While in the large $J_\perp$ limit, we can further reduce the bilayer model into the ESD model\cite{yang2023strong}, in which there are six states per site, with two bosonic states dubbed as d and b and four fermionic states. In the ESD model, as we increase the doping, we can see two different Fermi liquid phases with a Fermi surface volume jump by $1/2$ of the BZ per flavor. Ref.~{\onlinecite{yang2023strong}} focuses on the large $J_\perp$ limit, where the sFL and FL phases are confirmed at small and large doping $x$. At low temperature, the superconducting phase is found in both the sFL and FL phase, with a superconducting dome near $x\sim0.5$, which is attributed to the BCS to BEC to BCS crossover. In this paper, we focus on the region $x<0.5$, and by tuning $J_\perp$ we study the possible transition between the FL and sFL.

\subsection{sFL phase in the large $J_\perp $ limit}

We consider the $U_0, V_0\gg t$ limit and can then restrict to Eq.~\ref{eq:4+4_tJ_model} or Eq.~\ref{eq:4+6_t_J_model} depending on whether $\delta U$ is large or not. In either case, we expect two different metallic phases in the small and large $J_\perp$ regime at filling $n_t=n_b=1-x$ with small $x$.  When $J_\perp$ is small, one can expect a conventional Fermi liquid phase with Fermi surface volume $A_{FS}=\frac{1-x}{2}$ per flavor.

In the large $J_\perp$ limit, the doublon state is dominated by the spin-singlet. We can label this S=0 doublon as $\ket{b}=b^\dagger_i \ket{0}$ with $\ket{0}$ as the empty state. Together with four singlon states $\ket{a\sigma}=f^\dagger_{i;a\sigma}\ket{0}$, we reach a SU(4) t-J model with $4+1=5$ states per site. The t-J model is written as:

\begin{equation}
    \label{eq:4+1_tJ_model}
    \begin{aligned}
    H = -t\sum_{\langle ij \rangle}\sum_\alpha P_{4+1}c^\dagger_{i;\alpha} c_{j;\alpha}P_{4+1} 
    +...,
    \end{aligned}
\end{equation}

We can then do the standard slave boson theory\cite{lee2006doping} for this usual t-J model: $c_{i;a\sigma} = (-1)^{\eta_\sigma}\frac{1}{\sqrt{2}}f^\dagger_{\Bar{a}\Bar{\sigma}}b_i$ where $\eta_{\uparrow} = -\eta_{\downarrow} = 1$. We have the density $n_f=2x$ and $n_b=1-2x$. In terms of the parton construction, the model can be written as:

\begin{equation}
    H = \frac{t}{2}\sum_{\langle ij \rangle}\sum_{\alpha}f^\dagger_{j;\alpha} f_{i;\alpha}b^\dagger_i b_j.
\end{equation}

As usual, we can describe a Fermi liquid with condensation of the slave boson: $\langle b \rangle \neq 0$. Because $n_f=2x$ and each of the four flavor has filling $\frac{x}{2}$, we reach a Fermi surface volume of $A_{FS}=-\frac{x}{2}$, where the minus sign indicates that we have hole pocket because $c_i \sim f_i^\dagger$ and $f_i$ should be interpreted as annihilation of the hole.  Also the small Fermi surface should center at $\mathbf k=(\pi,\pi)$ due to the negative hopping of $f$. One can see now we have a different Fermi liquid in the infinite $J_\perp$ limit with Fermi surface volume smaller than the non-interacting limit by 1/2 of the BZ per flavor. We dub this phase as the sFL phase\cite{yang2023strong}. Our goal is to formulate a theory for the potential large to small Fermi surface transition by tuning $J_\perp$ at a fixed $x$.

\section{Deconfined Fermi liquid to Fermi liquid transition in the large $\delta U$ limit}

We first formulate a critical theory for the  FL to sFL transition in the large $\delta U$ limit, so we can restrict to the model in Eq.~\ref{eq:4+4_tJ_model}.  To capture the transition, we first need a unified framework to describe both the FL and the sFL phase. This can be done by a parton construction with a U(1)$\times$ U(1) gauge structure.

\subsection{Parton construction}

We introduce the standard Abrikosov fermion to represent the 4 singlon states: $f^\dagger_{i;a\sigma}\ket{0}$ with $a=t,b$. For the 4 doublon states, we introduce another fermion $\psi$ and $\psi^\dagger_{i;t\sigma}\psi^\dagger_{i;b\sigma^\prime}\ket{0}$ are 4 doublon states with $\sigma,\sigma^\prime = \uparrow,\downarrow$. Here $\psi_{i;a\sigma}$ annihilate a fermion at layer $a=t,b$ with spin $\sigma=\uparrow,\downarrow$ just as $f_{i;a\sigma}$. The electron operator projected to this Hilbert space is 

\begin{equation}
    \label{eq:electron_operator}
    c_{i;a\sigma} = \sum_{\sigma^\prime}f^\dagger_{i;\Bar{a}\sigma^\prime}\psi_{i;\Bar{a}\sigma^\prime}\psi_{i;a\sigma}.
\end{equation}
where $\Bar{a}$ is the opposite layer of $a$.

We have two local constraints: (I) $n_{i;f} + \frac{1}{2}\left( n_{i;\psi_t}+n_{i;\psi_b} \right)=1$ and (II) $n_{i;\psi_t}=n_{i;\psi_b}$ on each site. They generate two internal U(1) gauge fields $a_\mu$ and $b_\mu$ whose time components impose these two constraints as lagrangian multipliers. On average, we have density $n_f=2x$ while $n_{\psi_t} = n_{\psi_b} = 1-2x$. The total density of electrons is $n_f + n_{\psi_t}+n_{\psi_b}=2(1-x)=n_T$.

The physical electron operator Eq.~(\ref{eq:electron_operator}) is invariant under the following two internal U(1) gauge transformations: (1) $\psi_i \rightarrow \psi_i e^{i\theta_a (i)}$ and $f_i \rightarrow f_i e^{i2\theta_a (i)}$ for the U(1) gauge field $a_\mu$ (the subscript $a$ stands for the gauge field, not the layer index); (2) $f_i \rightarrow f_i$, $\psi_{i;t} \rightarrow \psi_{i;t} e^{i\theta_b (i)}$, $\psi_{i;b} \rightarrow \psi_{i;b}e^{-i\theta_b (i)}$ for the U(1) gauge field $b_\mu$. Meanwhile there is a global U(1) symmetry transformation: $c_{i;a}\rightarrow c_{i;a}e^{i\theta_c (i)}$. We can assign the charge to $\psi$, so under this global U(1) transformation, $f\rightarrow f$,$\psi_{i;a}\rightarrow \psi_{i;a}e^{i\frac{1}{2}\theta_c (i)}$. We introduce a probing field $A_\mu$ for this U(1) global symmetry. Another global U(1) symmetry transformation corresponding to the layer polarization $P_z$ is: $c_{i;t}\rightarrow c_{i;t}e^{i\theta_d (i)}$, $c_{i;b}\rightarrow c_{i;b}e^{-i\theta_d (i)}$. We assign the charge to $f$, so under this global U(1) transformation, $\psi \rightarrow \psi$, $f_{i;t} \rightarrow f_{i;t}e^{i\theta_d (i)}$, $f_{i;b}\rightarrow f_{i;b}e^{-i\theta_d (i)}$. We also introduce a probing field $B_\mu$ for the layer U(1). In the end, we have $f_t$ couples to $2a_\mu+B_\mu$, $f_b$ couples to $2a_\mu - B_\mu$, $\psi_t$ couples to $\frac{1}{2}A_\mu + a_\mu + b_\mu$ and $\psi_b$ couples to $\frac{1}{2}A_\mu + a_\mu-b_\mu$. The gauge charge of each field is summarized in  Table.~\ref{gauge_charge}

\begin{table}[h!]
\centering
\begin{tabular}{ |c|c|c|c|c|}
\hline
  & $f_{t\sigma}$ & $f_{b\sigma}$& $\psi_{t\sigma}$&$\psi_{b\sigma}$ \\ 
  \hline
 $a_\mu$& +2&+2&+1&+1\\
 \hline
 $b_\mu$& 0&0&+1&-1\\
 \hline
 $A_\mu$&0&0&+$\frac{1}{2}$&+$\frac{1}{2}$\\
 \hline
 $B_\mu$&+1&-1&0&0\\
 \hline
\end{tabular}
\caption{Gauge fields and the corresponding charges for each operator. $a_\mu$ and $b_\mu$ are the gauge field introduced by the local constraints, while $A_\mu$ and $B_\mu$ are the probing field related to the global $U(1)$ symmetry in each layer.}
\label{gauge_charge}
\end{table}

Rewriting the Hamiltonian Eq.~(\ref{eq:4+4_tJ_model}) in terms of the partons and doing the mean-field decoupling, we can obtain the following two different possible mean field ansatz: 

\begin{equation}
\label{eq:definition_Phi_Delta}
\begin{aligned}
    &H^{(1)}_{\mathrm{MF}} = -\sum_i\sum_{a=t,b}\Phi_a f^\dagger_{i;a\sigma}\psi_{i;a\sigma} + \mathrm{h.c.}, \\
    &H^{(2)}_{\mathrm{MF}} = -\Delta \sum_i \epsilon_{\sigma\sigma^\prime} \psi_{i;t\sigma}\psi_{i;b\sigma^\prime} + \mathrm{h.c.}.
\end{aligned}
\end{equation}

\begin{figure}
    \centering
    \includegraphics[scale=0.5]{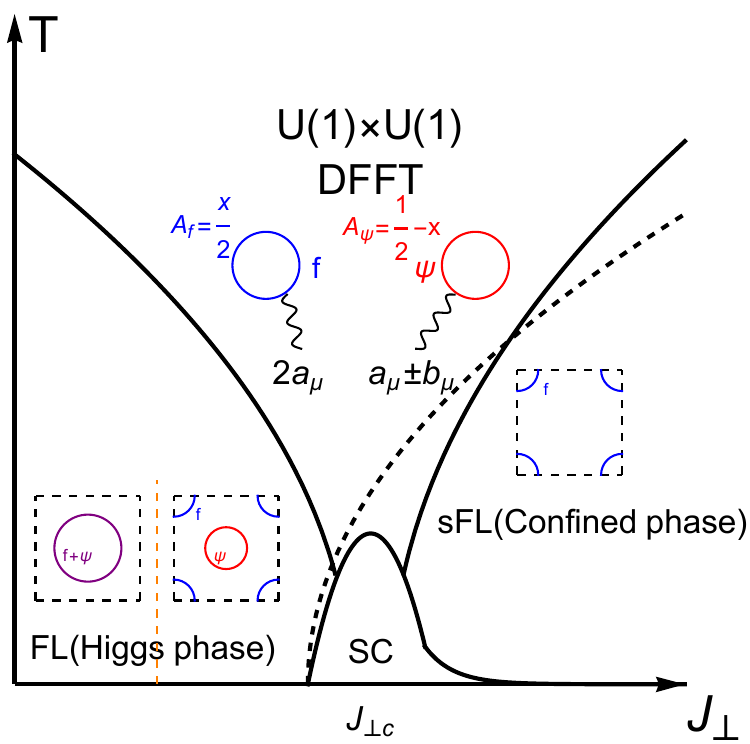}
    \caption{A schematic phase diagram with $J_\perp$, which tunes the mass of $\Phi_a$ and the temperature $T$ in the large $\delta U$ regime. In the FL phase, $\langle \Phi_a \rangle \neq 0$ and $\Delta = 0$. When $\langle \Phi_a \rangle$ is large, $f$ and $\psi$ share a Fermi surface while when $\langle \Phi_a \rangle$ is small, $\psi$ has a Fermi surface around $\Gamma$ point and $f$ has a Fermi surface around $(\pi,\pi)$. In the sFL phase, $\langle \Phi_a \rangle=0$ and $\Delta \neq 0$. $\psi$ now is gapped while $f$ forms a small hole pocket. The two crossover boundaries are the usual V-shape quantum critical region for the critical boson $\Phi$. The dashed curve denote the pairing of $\psi$. The yellow dashed line denotes a Lifshitz transition within the FL phase. At low temperature, there is a superconductor dome due to the pairing instability of the DFFT critical point.  Above the pairing scale in the critical regime, we have two types of Fermi surfaces from $f$ and $\psi$ particles. Both of them couple to internal gauge field $a_\mu, b_\mu$ and the electron is a three particle bound state $c\sim f^\dagger \psi \psi$ and thus there is no coherent single electron quasiparticle.}
    \label{fig:4+4_phase_diagram}
\end{figure}

The variational parameters $\Phi_a$ and $\Delta$ need to be decided by optimizing the energy at each fixed $J_\perp$. Ideally this should be done through the variational Monte carlo (VMC) calculation because the simple self-consistent mean field analysis is known to be not trustable due to the constraint.  Because our focus here is the universal theory of the transition, we leave the VMC calculation with detailed energetical analysis to future. Here we simply point out two different phases accessible in this parton framework:

\begin{itemize}
    \item \textbf{FL phase: $\Phi_a\neq 0, \Delta=0$}.  Now $f_a$ and $\psi_a$ hybridize and both of them can be identified as the electron operator $c_{a\sigma}\sim f_{a\sigma} \sim \psi_{a\sigma}$. The total density is $n_T=2-2x$ and the total Fermi surface volume per flavor is $\frac{1-x}{2}$ because we have four identical Fermi surfaces from the layer and spin.  For the gauge field, $a_\mu$ is higgsed to be locked to $\frac{1}{2}A_\mu$ while $b_\mu$ is locked to $B_\mu$. This is a higgsed phase of the U(1)$\times$ U(1) gauge theory.
    \item \textbf{sFL phase: $\Phi_a=0, \Delta \neq 0$} Now $\psi_a$ is gapped out because of the pairing. $a_\mu$ is higgsed to be locked to $-\frac{1}{2} A_\mu$ while $b_\mu$ is confined. In the end we have $c_{a\sigma} \sim f^\dagger_{\bar a \bar \sigma}$. $f$ couples to $-A_\mu$ and should be interpreted as the hole operator. Because $n_f=2x$, we expect Fermi surface volume $A_{FS}=-\frac{x}{2}$ per flavor.
\end{itemize}

Now we see that we can capture both the FL and the sFL phase within one framework. Although our formalism introduces $U(1) \times U(1)$ gauge field, the FL and sFL phases are conventional in the sense that the emergent gauge field is either higgsed or confined. Following our argument before, we naturally expect that the above two ansatz correspond to the small $J_\perp$ and the large $J_\perp$ regime. The next question is about the intermediate regime. In principle we can have an intermediate phase from the other two ansatz: (III) A superconductor phase with $\Phi_a\neq 0, \Delta \neq 0$ and (IV) A deconfined metal (DM) phase with $\Phi_a=0,\Delta=0$.  Another possibility is a first order transition between the FL and the sFL phase directly. The most interesting possibility is a continuous direct transition from FL to sFL.  If we start from the FL phase with finite $\Phi_a$, the question is whether the onset of the pairing $\Delta$ can coincide with the disappearance of the higgs condensation $\Phi_a$. Depending on whether this happens and assuming no superconductivity, there should be a critical point or a line separating the superconductor phase from the DM phase in between FL and sFL. In the mean field level this is impossible without fine tuning, but the gauge fluctuation can change the story completely. One famous example is the Neel to VBS DQCP\cite{senthil2004deconfined} where the confinement happens immediately after the higgs phase is destroyed due to the proliferation of the monopole.  In our case we will show a similar scenario with the onsets of the pairing driven by the destruction of the higgs condensation $\Phi_a$.

\subsection{Critical theory}

We start from the FL phase with $\langle \Phi_a \rangle \neq 0$, then we expect $\Phi_a$ decrease with $J_\perp$ because eventually we have the sFL phase with $\langle \Phi_a \rangle =0$ at large $J_\perp$. Next we will formulate a critical theory associated with the Higgs transition of $\Phi_a$ which vanishes at a critical value $J_\perp^c$.  This critical theory can be described by the following Lagrangian:

\begin{equation}
    \label{eq:Lc}
    \mathcal{L}_{\mathrm{c}} = \mathcal{L}_{\Phi} + \mathcal{L}_f + \mathcal{L}_\psi,
\end{equation}
where $\mathcal{L}_\Phi$ is

\begin{widetext}
    \begin{equation}
        \begin{aligned}
            \mathcal{L}_\Phi &= \Bar{\Phi}_t \left( \partial_\tau - i\left( -\frac{A_0}{2}+B_0+a_0-b_0 \right) \right)\Phi_t + \frac{1}{2m_{\Phi}}\Bar{\Phi}_t\left( -i\vec\nabla-\left( -\frac{\vec{A}}{2}+\vec{B}+\vec{a}-\vec{b} \right) \right)^2 \Phi_t \\
        &+\Bar{\Phi}_b \left( \partial_\tau - i\left( -\frac{A_0}{2}-B_0+a_0+b_0 \right) \right)\Phi_b + \frac{1}{2m_{\Phi}}\Bar{\Phi}_b\left( -i\vec\nabla-\left( -\frac{\vec{A}}{2}-\vec{B}+\vec{a}+\vec{b} \right) \right)^2 \Phi_b \\
        &-\mu_\Phi \left( \left| \Phi_t \right|^2 + \left| \Phi_b \right|^2 \right) + \lambda_1 \left( \left| \Phi_t \right|^2 + \left| \Phi_b \right|^2 \right)^2 + \lambda_2 \left| \Phi_t \right|^2 \left| \Phi_b \right|^2. \\
        \end{aligned}
    \end{equation}
\end{widetext}

Here we assume that there is always a mirror reflection symmetry $\mathcal M$ which exchanges the two layers, so $\Phi_a$ and $\Phi_b$ have the same mass $\mu_\Phi$. $\mathcal L_\Phi$ is just the standard action for the Higgs transition. Here the boson fields $\Phi_a, \Phi_b$ couples to the internal U(1) gauge fields $a_\mu, b_\mu$ and also two probing fields $A_\mu, B_\mu$. When $\mu_\Phi>0$, we have $\langle \Phi_a \rangle=\langle \Phi_b \rangle \neq 0$, which locks $a_\mu=\frac{1}{2} A_\mu$ and $b_\mu=B_\mu$.  When $\mu_\Phi<0$, $\Phi_a$ is gapped and the two internal U(1) gauge fields $a_\mu, b_\mu$ become alive.  

$\mathcal L_f$ and $\mathcal L_\psi$ contain the action of the fermion $f$ and $\psi$:

\begin{widetext}
\begin{equation}
    \begin{aligned}
        \mathcal{L}_f &= \Bar{f}_{t;\sigma}\left( \partial_\tau - i\left( 2a_0+B_0 \right) \right)f_{t;\sigma} +\frac{1}{2m_f}\Bar{f}_{t;\sigma}\left( -i\vec\nabla - \left( 2\vec{a}+\vec{B} \right) \right)^2 f_{t;\sigma}   \\
        &+ \Bar{f}_{b;\sigma}\left( \partial_\tau - i\left( 2a_0-B_0 \right) \right)f_{b;\sigma} +\frac{1}{2m_f}\Bar{f}_{b;\sigma}\left( -i\vec\nabla - \left( 2\vec{a}-\vec{B} \right) \right)^2 f_{b;\sigma}  \\
        &-\mu_f \sum_{a=t,b}\Bar{f}_{a;\sigma}f_{a;\sigma}, \\
        \mathcal{L}_\psi&= \Bar{\psi}_{t;\sigma}\left( \partial_\tau - i\left( \frac{A_0}{2}+a_0+b_0 \right) \right)\psi_{t;\sigma} + \frac{1}{2m_{\psi}}\Bar{\psi}_{t;\sigma} \left( -i\vec\nabla - \left( \frac{\vec{A}}{2}+\vec{a}+\vec{b} \right) \right)^2 \psi_{t;\sigma} \\
        &+\Bar{\psi}_{b;\sigma}\left( \partial_\tau - i\left( \frac{A_0}{2}+a_0-b_0 \right) \right)\psi_{b;\sigma} + \frac{1}{2m_{\psi}}\Bar{\psi}_{b;\sigma} \left( -i\vec\nabla - \left( \frac{\vec{A}}{2}+\vec{a}-\vec{b} \right) \right)^2 \psi_{b;\sigma}  \\
        &-\mu_{\psi}\sum_{a=t,b}\Bar{\psi}_{a;\sigma}\psi_{a;\sigma}.
    \end{aligned}
\end{equation}
\end{widetext}
Remember that $f_a$ couples to $2a_\mu\pm B_\mu$ for $a=t,b$ and  $\psi_a$ couples to $\frac{1}{2}A_\mu + a_\mu \pm b_\mu$ for $a=t,b$.  In the higgs phase with $\langle \Phi_a \rangle \neq 0$, both $f_{a\sigma}, \psi_{a\sigma}$ can be identified as electron operator.  When $|\Phi_a|$ is large enough, $f,\psi$ hybridize together to form a single large Fermi surface.  Then at small but finite $\Phi_a$, we expect separate Fermi surfaces dominated by $f$ and $\psi$.  But their total Fermi surface volume is still $A_{FS}=\frac{1-x}{2}$ per flavor and it is still a conventional FL phase. When approaching the critical point $\mu_\Phi=0$, the quasi particle residue of both Fermi surfaces vanish.  At the critical point, the Fermi surface volumes per flavor from $f$ and $\psi$ are $\frac{x}{2}$ and $\frac{1}{2}-x$ respectively.   In principle that there should be a Yukawa coupling $\delta L=g \Phi f^\dagger_{a\sigma} \psi_{a\sigma}$. But given the mismatch of the Fermi surfaces from $f$ and $\psi$ in the momentum space, this coupling is irrelevant because the critical boson $\Phi$ is mainly at zero momentum.

When $\mu_\Phi<0$, $\Phi_a$ is gapped and we can ignore $\mathcal L_\Phi$. But now the two internal U(1) gauge fields $a_\mu, b_\mu$ become alive and we need to decide whether the Fermi surfaces from $f$ and $\psi$ are stable to the gauge fluctuation or not. $f$ only couples to $2a_\mu$ and  the physics is then similar to the familiar U(1) spin liquid with spinon Fermi surface. From the previous works we know that the Fermi surface from $f$ should be stable.  On the other hand, $\psi_t$ and $\psi_b$ couple to $a_\mu$ with the same charge, but couple to $b_\mu$ with opposite charge. It is known that $b_\mu$ will mediate attractive interaction between $\psi_t$ and $\psi_b$\cite{zhang2020deconfined}. We will show next that this attractive interaction is stronger than the repulsive interaction from $a_\mu$, leading to a pairing instability and an intermediate superconductor phase between the two Fermi liquids.

\subsection{Pairing instability and superconductor dome}

Here we analyze the pairing instability for $\mu_\Phi\leq 0$. First, the fermion bubble diagrams give rise to the following effective photon action:

\begin{equation}
    \begin{aligned}
        \mathcal{L}_{a,b} &= \frac{1}{2}\left( \frac{1}{e_{a,0}^2}\left| \mathbf{q} \right|^2 + \kappa_a \frac{|\omega|}{|\mathbf{q}|} \right)\left| a(\omega,\mathbf{q}) \right|^2\\
        &+ \frac{1}{2}\left( \frac{1}{e_{b,0}^2}\left| \mathbf{q} \right|^2 + \kappa_b \frac{|\omega|}{|\mathbf{q}|} \right)\left| b(\omega,\mathbf{q}) \right|^2,
    \end{aligned}
\end{equation}
with

\begin{equation}
    \begin{aligned}
        \frac{1}{e_{a,0}^2} &= \frac{1}{6\pi m_\psi} + \frac{2}{3\pi m_f}, \\
        \frac{1}{e_{b,0}^2} &= \frac{1}{6\pi m_\psi}.
    \end{aligned}
\end{equation}
Here $m_f$ and $m_\psi$ are effective mass for $f$ and $\psi$ fermions. We note that $\Phi_a$ does not contribute to the photon action when $\mu_\Phi \leq 0$.

Since $\psi_{t/b}$ couples to $a_\mu\pm b_\mu$, the exchange of the $a_\mu$ ( $b_\mu$) photon induces repulsive (attractive) interaction between the two  layers for the $\psi$ fermion. Thus there is the possibility of pairing instability, The renormalization group (RG) flow equation of the interaction strength $V$ in the inter-layer Cooper channel (at any angular momentum) is \cite{Metlitski2015}:

\begin{equation}
    \frac{d \Tilde{V}}{d l} = \left( \alpha_{a,\psi} - \alpha_{b,\psi} \right) - \Tilde{V}^2,
\end{equation}
where $\alpha_{a/b,\psi} = \frac{e_{a/b}^2 v_{F,\psi}}{4\pi^2}$ is the coupling strength, where $v_{F,\psi}$ is the Fermi velocity of $\psi$. Here $l$ is the RG step. The first term comes from the exchange of photons, while the second term is the usual BCS flow for Fermi liquids. In Appendix.~\ref{sec:critical_theory}, we also show that $\alpha_{a,\psi}/\alpha_{b,\psi}$ does not  flow\cite{zou2020deconfined,zhang2020deconfined}. In our case we have $e_{a,0}^2 < e_{b,0}^2$, so we have $\alpha_{a;\psi}-\alpha_{b;\psi}<0$ and the interaction $\tilde V$ flows to $-\infty$ even if the initial interaction is repulsive. Assuming the initial $\Tilde{V}$ is positive and large, we estimate the superconducting gap to be $\Delta \sim \Lambda_\omega \exp (-l_p)$, where 

\begin{equation}
    l_p \approx \pi/\sqrt{\alpha_{b,\psi}^0 - \alpha_{a,\psi}^0} = \sqrt{\frac{\pi^3 \left( m_f + 4m_\psi \right)}{6v_{F,\psi}m_\psi^2}}
\end{equation}
in the calculation with $\epsilon$ expansion\cite{Metlitski2015}. $\Lambda_\omega$ is the energy cutoff in RG.  Note that the pairing scale is quite small if $m_f$ is large. Generically the pairing scale is smaller than that from the nematic critical point\cite{Metlitski2015}, where the critical boson induces strong attractive interaction. For our case, because we also have the balance from the repulsive interaction from the other gauge field $a_\mu$, we expect suppressed pairing scale and there should be a large critical regime  at temperature above the pairing energy scale.

Note that the above analysis holds for $\mu_\Phi \leq 0$. Even if $\mu_\Phi=0$, the gapless higgs boson $\Phi_a$ does not alter the conclusion because the higgs boson does not contribute to the photon self energy. This means that the pairing instability exists already at the critical point. So even at $\mu_\Phi=0$, there should be a finite pairing term $\Delta \epsilon_{\sigma \sigma'} \psi_{t;\sigma} \psi_{b;\sigma'}$ with $\Delta \neq 0$ at zero temperature. So we expect the onsets of $\Delta$ must happen before the disappearance of $\Phi_a$, as is illustrated by the black dashed line in Fig.~\ref{fig:4+4_phase_diagram}.  In the intermediate region, $\Phi_a$ and $\Delta$ coexist. When there is finite $\Phi_a$, we expect $c_{a\sigma}\sim \psi_{a\sigma}\sim f_{a\sigma}$. The pairing of $\psi$ means pairing of electron and this is a superconductor phase with inter-layer pairing. Note that due to finite $\Phi$, pairing of $\psi$ transmits to pairing of $f$ and all of the Fermi surfaces should be gapped.  The pairing instability happens at any angular momentum channel. Microscopic details are needed to decide which angular momentum wins. One natural guess is a $s'$-wave inter-layer pairing to avoid on-site inter-layer repulsion.

When $\mu_\Phi<0$, $\Phi_a$ is gapped. Now the pairing of $\psi$ does not mean pairing of electron anymore. Actually we simply have $c_{a\sigma}\sim f^\dagger_{a\sigma}$ and we have a small hole pocket from $f$, while the fermi surface from $\psi$ is gapped. This is the sFL phase. Now the gauge field $a_\mu$ is higgsed by the non-zero $\Delta$. The U(1) gauge field $b_\mu$ does not couple to any gapless matter field and will be confined due to the proliferation of the monopole of $b_\mu$ in 2+1d\cite{polyakov1987gauge}. As we know the sFL phase is allowed by the Luttinger theory, the monopole proliferation does not need to break any symmetry. We note that the sFL phase may still have a weak pairing instability at very low temperature.  Actually now the term $\Phi_a f^\dagger_{a\sigma} \psi_{a\sigma}$ leads to a term $\delta \mathcal L=g' \Phi_t \Phi_b 
 \epsilon_{\sigma \sigma'} f^\dagger_{t\sigma} f^\dagger_{b\sigma'}$ through a second-order process, given that $\epsilon_{\sigma \sigma'} \langle \psi_{t\sigma} \psi_{b\sigma'} \rangle \neq 0$. Now the small hole pocket from $f$ couples to the composite boson field $\Phi=\Phi_t \Phi_b$ in the form of a boson-fermion model. Note that $\Phi_t, \Phi_b$ couples to $\pm b_\mu$. The confinement of $b_\mu$ means that they now strongly bound to each other and we can treat $\Phi=\Phi_t \Phi_b$ as a well-defined particle. $\Phi$ is actually a virtual Cooper pair now with an energy cost $2 |\mu_\Phi|$. The exchange of the virtual Cooper pair leads to an attractive interaction for the $f$ fermion with $V\sim - \frac{g^{'2}}{|\mu_\Phi|}$. Hence the small hole pocket in the sFL phase has a BCS instability at low temperature.  A similar mechanism of pairing instability of the sFL phase from virtual Cooper pair has been discussed in our previous work\cite{yang2023strong}, but there the virtual Cooper is the on-site pair from $J_\perp$ term at finite repulsion $V$. In our current case the on-site inter-layer Cooper pair is pushed to infinite energy because we take $V=+\infty$ and the mechanism in our previous work does not apply anymore.  The virtual Cooper pair we discussed here is from the bound state of the Higgs boson $\Phi=\Phi_t \Phi_b$ and plays a role only not too far away from the critical regime.  Therefore this is a completely new mechanism of superconductivity associated with the deconfined FL to FL criticality.

 \subsection{Property of the critical regime}

 As illustrated in Fig.~\ref{fig:4+4_phase_diagram}, we expect a critical regime governed by the DFFT critical point at finite temperature above the superconductor dome.  In the critical regime we have $\langle \Phi_a \rangle= \Delta =0$. So now the two U(1) gauge fields $a_\mu, b_\mu$ are deconfined.  We have two types of Fermi surfaces from $f_{a\sigma}$ and $\psi_{a\sigma}$ for each flavor $a=t,b$ and $\sigma=\uparrow,\downarrow$. Their fermi surface volumes are fixed to be $A_f=\frac{x}{2}$ and $A_\psi=\frac{1}{2}-x$ per flavor when the total density per site (summed over layer and spin) is $n_T=2-2x$. In the critical regime described by the theory in Eq.~\ref{eq:Lc}, the microscopic electron operator is a three-particle bound state\cite{PhysRevB.78.045109,PhysRevLett.102.186401} of the $f$ and $\psi$ fermions:

\begin{equation}
    c_{a\sigma}(\tau,x)=\sum_{\sigma'}f^\dagger_{\bar a \sigma'}(\tau,x)\psi_{\bar a;\sigma'}(\tau,x) \psi_{a;\sigma}(\tau,x)
\end{equation}

\begin{figure}
    \centering
    \includegraphics[scale=0.28]{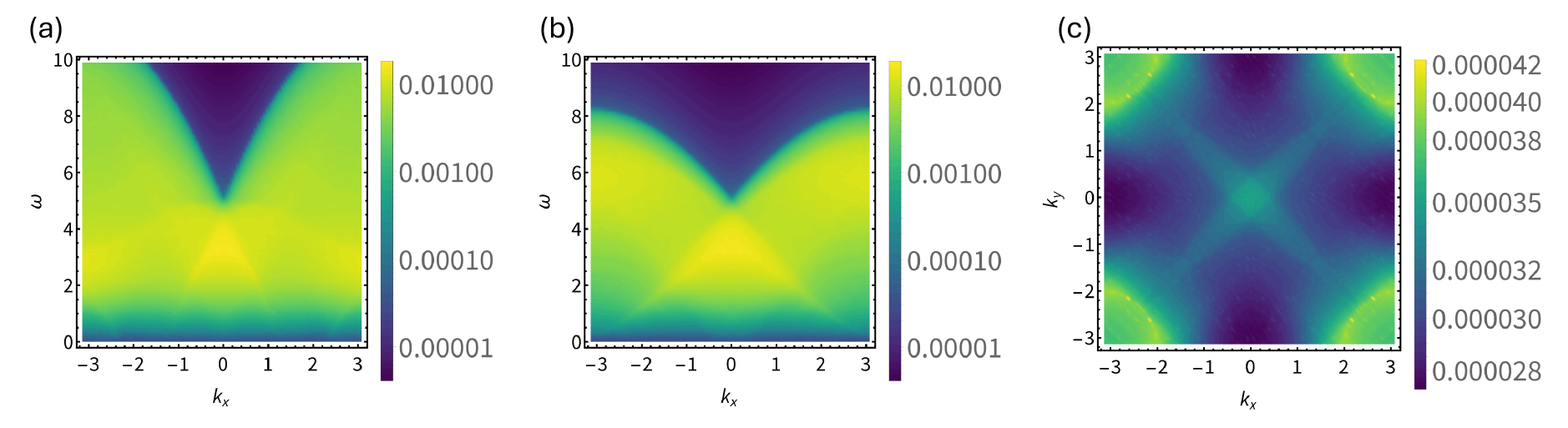}
    \caption{(a), (b), (c) the electron spectral function $A(k_x=k_y,\omega)$, $A(k_x,k_y = 0,\omega)$, $A(k_x, k_y, \omega = 0)$ 
 in the critical regime of the FL to sFL transition at temperature above the superconductor dome in Fig.~\ref{fig:4+4_phase_diagram}. In the calculation, we use the dispersion $E_f=2t(\cos k_x+\cos k_y)-\mu_f$ and $E_\psi=-2t(\cos k_x+\cos k_y)-\mu_\psi$, and the parameters are $t=1$, $\mu_f= -2.844$, $\mu_\psi = -1.057$ are chosen to make doping $x = 0.2$. In the critical region, the system is in the decofined phase, in which the electron operator is a three-particle bound state, so there is no coherence peak in the spectral function.}
    \label{fig:spectral_function}
\end{figure}

\begin{figure}
    \centering
    \includegraphics[scale=0.28]{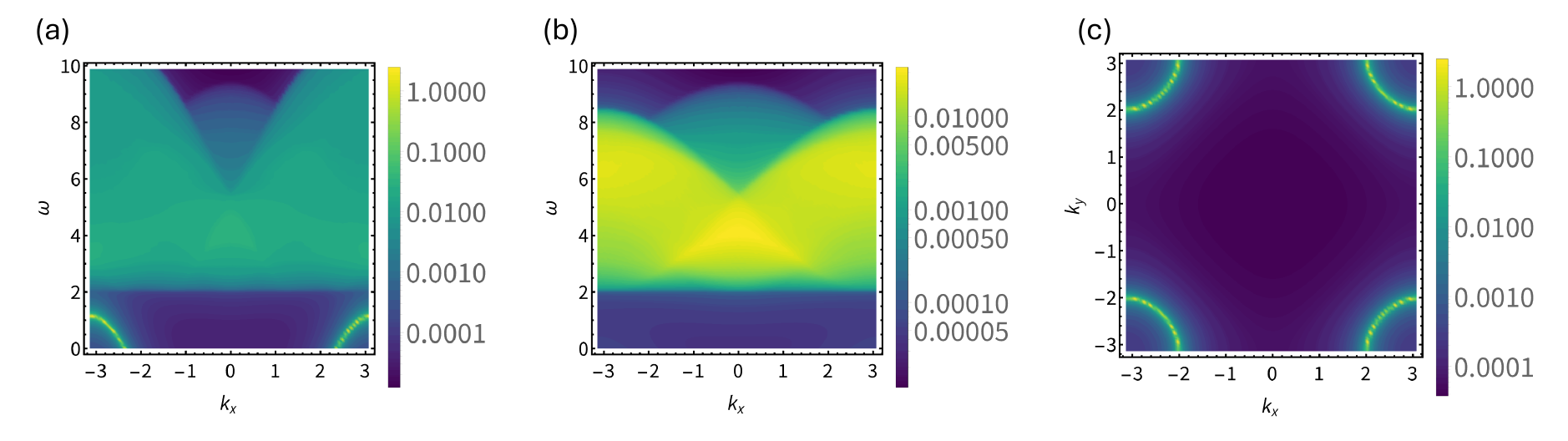}
    \caption{(a), (b), (c) the electron spectral function $A(k_x=k_y,\omega)$, $A(k_x,k_y = 0,\omega)$, $A(k_x, k_y, \omega = 0)$ in sFL. In addition to parameters we used in Fig.~\ref{fig:spectral_function}, we choose the pairing $\Delta = 1$. In the sFL phase, $\psi$ is gapped, so the electron operator is roughly $c\sim f^\dag$. We can see the Fermi surface of $f$ from the spectral function.}
    \label{fig:spectral_function_sFL}
\end{figure}

\begin{figure}
    \centering
    \includegraphics[scale=0.28]{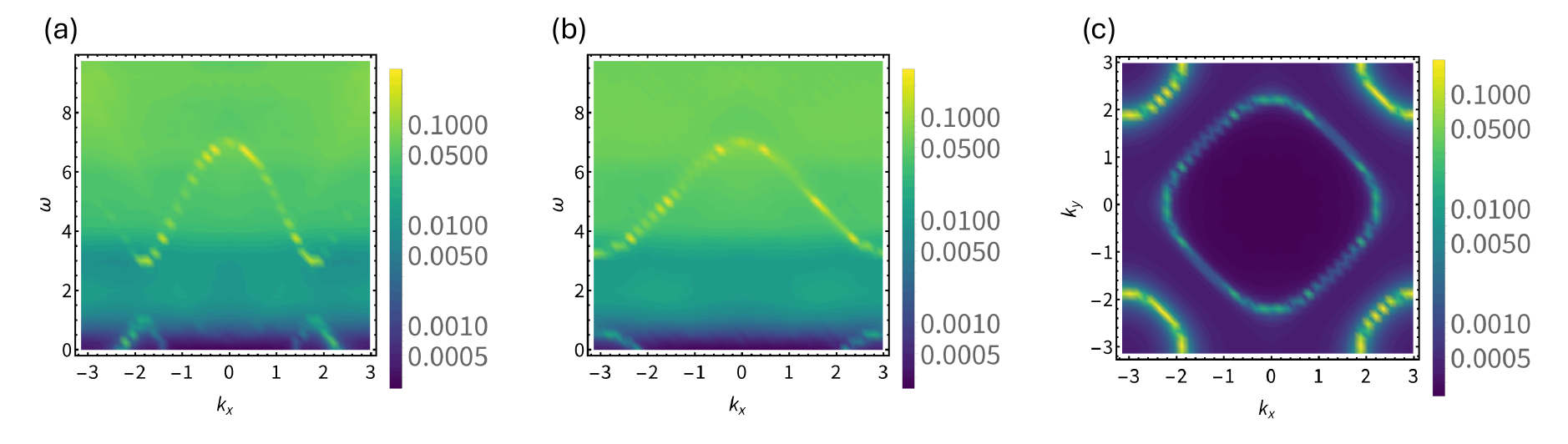}
    \caption{(a), (b), (c) the electron spectral function $A(k_x=k_y,\omega)$, $A(k_x,k_y = 0,\omega)$, $A(k_x, k_y, \omega = 0)$ in FL. In addition to parameters we used in Fig.~\ref{fig:spectral_function}, we choose the condensate $\Phi = 1$. In the FL phase, $\Phi\neq 0$, the electron operator is roughly $c\sim f^\dag\sim \psi$. We can see the Fermi surfaces of both $f$ and $\psi$ from the spectral function.}
    \label{fig:spectral_function_FL}
\end{figure}

The elementary particles $f,\psi$ in the low energy theory couple to $2a_\mu$ and $\frac{1}{2}A_\mu+a_\mu \pm b_\mu$ respectively. None of them is gauge invariant, so the Fermi surface of $f$ or $\psi$ is not detectable by physical probes. Instead the physical Green function now is

\begin{align}
    G^R_c(\tau,x)&=-\mathbbm{i}\Theta(\tau)\langle c_{a\sigma}(\tau,x) c^\dag_{a\sigma}(0,0)\rangle \notag \\ 
    &~\sim \langle f(\tau,x)\psi^\dagger_t(\tau,x) \psi^\dagger_b(\tau,x) f^\dagger(0,0)\psi_t(0,0)\psi_b(0,0)\rangle
\end{align}
where in the last line we suppress the spin index for simplicity. One obvious implication is that the physical Green function now has a large power law scaling dimension. In mean field level, it is three times larger than the usual free fermion from the Wick theorem. In $(\omega,k)$ space, $G_c(\omega,k)$ is from a complicated convolution and we do not expect any coherent quasi-particle peak in ARPES or STM probes even without considering gauge fluctuations. In Fig.~\ref{fig:4+4_phase_diagram}, we schematically show the properties in each phase. Here in Fig.~\ref{fig:spectral_function} to Fig.~\ref{fig:spectral_function_FL}, we calculate the electron spectral function $A({\bf k},\omega)=-\frac{1}{\pi}\text{Im}{G_c({\bf k},\omega)}$. We can see that in the critical region, there is no coherence peak, while in the sFL phase and the FL, we can see Fermi surface of quasiparticles, which is consistent with the schematical phase diagram.

In transport the system should behave as a metal. Under $A_\mu$, $f,\psi$ and $a_\mu, b_\mu$ will respond.  Let us apply a constant external electric field $\vec E=-\partial_t \vec A-\vec \nabla A_0$. We also define the internal electric field: $\vec e_a=-\partial_t \vec a-\vec \nabla a_0$ and $\vec e_b=-\partial_t \vec b-\vec \nabla b_0$. 

Then the usual conductivity relations of $f$ and $\psi$ give us:

\begin{align}
    \vec J_f&=\sigma_f (2 \vec e_a) \notag \\
    \vec J_{\psi_t}&=\frac{1}{2}\sigma_\psi (\frac{1}{2} \vec E+\vec e_a+\vec e_b) \notag \\
     \vec J_{\psi_b}&=\frac{1}{2}\sigma_\psi (\frac{1}{2} \vec E+\vec e_a-\vec e_b)
\end{align}
where $\sigma_f$ is the conductivity of the $f$ fermion summed over the two layers. $\sigma_\psi$ is  the conductivity of the $\psi$ fermion summed over  the two layers. Here we assume a layer exchange symmetry. 

From the above three equations we can eliminate $\vec e_a, \vec e_b$ to reach $-\frac{\vec J_f}{\sigma_f}+\frac{2\vec J_{\psi_t}+2\vec J_{\psi_b}}{\sigma_\psi}=\vec E$. From the local constraint $n_{i;f}+\frac{1}{2} (n_{i;\psi_t}+n_{i;\psi_b})=1$ and $n_{i;\psi_t}=n_{i;\psi_b}$ we have $-\vec J_f=\vec J_{\psi_t}=\vec J_{\psi_b}=\vec J_c$ where $\vec J_c$ is the physical current.  In the end we get $\vec J_c (\frac{1}{\sigma_f}+\frac{4}{\sigma_\psi})=\vec E$.  Finally we reach the Ioffe-Larkin rule for the resistivity:

\begin{equation}
    \rho_c=\rho_f+4\rho_\psi
\end{equation}

$\rho_f$ and $\rho_\psi$ are resistivities of the $f$ and $\psi$ Fermi surfaces. We expect them to be metallic in the sense that they increase with temperature $T$. However, the exact behavior of $\rho_f$ and $\rho_\psi$ are complicated due to the coupling to the internal U(1) gauge fields\cite{lee2018recent}.  We leave the future work to decide the transport behavior of the DFFT critical regime.  Another interesting question is the low energy emergent symmetry and anomaly of the DFFT critical line. Recently it was shown that some non-Fermi liquid share the same emergent symmetry and anomaly structure as the Fermi liquid and they all belong to the so called ersatz Fermi liquid (EFL)\cite{else2021non}.  Our DFFT critical regime is apparently compressible, so one can ask similar question. We conjecture that it does not belong to the earsatz Fermi liquid and needs a different description in terms of emergent symmetry and anomaly, which we leave to future work.

\section{Suppression of pairing by $\delta U$ and deconfined metallic phase}

In the large $\delta U$ regime, we have shown that there must be a superconductor dome between the FL and sFL phase at zero temperature. The DFFT criticality can only be revealed at finite temperature. It is then interesting to ask whether we can fully suppress the pairing instability. This turns out to be possible by decreasing $\delta U$. We will study the fate of the critical regime with $\mu_\Phi$ and $\delta U$ to be the two relevant directions. $\mu_\Phi$ governs the transition between sFL and FL, while $\delta U$ tunes another Higgs transition and suppresses the pairing instability.

Now let us start from the full model in  Eq.~(\ref{eq:general_SU4_model}). We still take $U$ to be large, but treat $\delta U$ as a tuning parameter. The restricted Hilbert space now has 4 singlon states and 6 doublon states (see Fig.~(\ref{fig:4+6_hilbert_space})) and the Hamiltonian is Eq.~(\ref{eq:4+6_t_J_model}). We still use similar parton construction with $f^\dagger_{i;a\sigma}$ to create singlon states and $\psi^\dagger_{i;a\sigma}$ to create doublon states. The difference now is that we have two extra doublon states: $\psi^\dagger_{i;t\uparrow}\psi^\dagger_{i;t\downarrow}\ket{0}$ and $\psi^\dagger_{i;b\uparrow}\psi^\dagger_{i;b\downarrow}\ket{0}$ at each site. In this case, there are two tuning parameters $\delta U$ and $J_\perp$ in the microscopic model.

At each fixed $\delta U$, we still expect the FL and sFL phase in the small and large $J_\perp$ limit. In our low energy critical theory, tuning $J_\perp$ still effectively changes the mass $\mu_\Phi$ to gap out the higgs condensation $\Phi_a$. We will show that tuning $\delta U$ changes the mass of another boson $\varphi$, which higgses the gauge field $b_\mu$. By tuning both parameters we can approach a deconfined critical point, which we argue is stable to pairing and may survive down to zero temperature.

\begin{figure}
    \centering
    \includegraphics[scale=0.6]{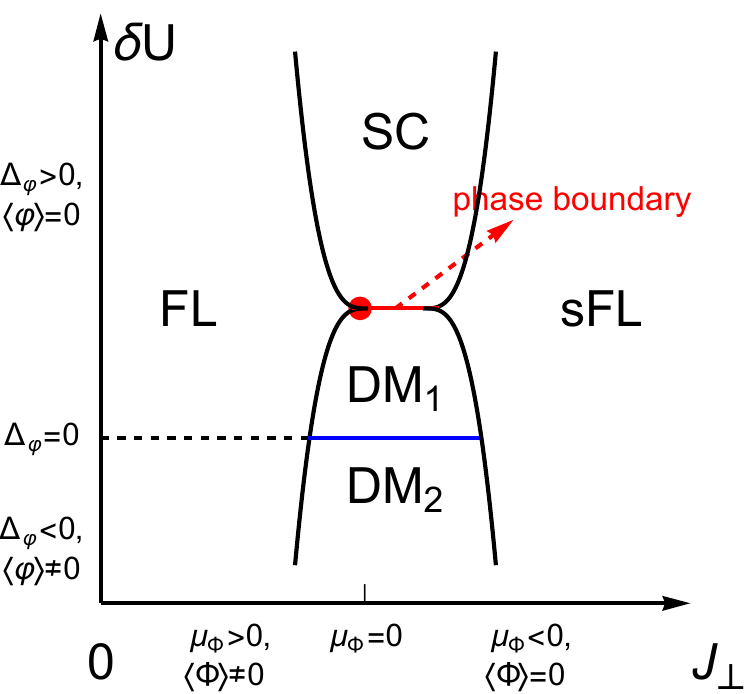}
    \caption{A schematic phase diagram at zero temperature with $J_\perp$ and $\delta U$ as the tuning parameters, which effectively tune the mass of $\Phi$ ($\mu_\Phi$) and $\varphi$ ($\Delta_\varphi$). $\mu_\Phi$ and $\Delta_\varphi$ are two relevant directions. There is a critical value $\Delta_\varphi = \Delta_{\varphi;c} = m_f^2 / 16$ tuned by $\delta U$, which separates the superconducting phase and a stable intermediate deconfined metal (DM) phase at $\mu_\Phi=0$ tuned by $J_\perp$. When $\Delta_\varphi$ is smaller than $\Delta_{\varphi;c}$, there are two deconfined metal(DM) phases. In DM1 there are two deconfined gauge fields $a_\mu$ and $b_\mu$ while in DM2 $b_\mu$ is higgsed by $\langle \varphi \rangle \neq 0$. }
    \label{fig:phase_diagram}
\end{figure}

The role of $\delta U$ is to add an energy penalty to the last two doublon states which violates the condition $n_{i;\psi_t}=n_{i;\psi_b}$: $H'=\delta U\sum_i (n_{i;\psi_t}-n_{i;\psi_b})^2$. Now the previous local constraint $n_{i;\psi_t}=n_{i;\psi_b}$ is not exact anymore unless $\delta U=+\infty$. As the U(1) gauge field $b_\mu$ originates from this constraint, we expect that $b_\mu$ is alive at the large $\delta U$ regime, but should disappear in the small $\delta U$ regime.  How do we capture this evolution?  The best way is to introduce another slave rotor corresponding to 

\begin{equation}
    L_i = \frac{1}{2}\left( n_{i,\psi_t} - n_{i,\psi_b} \right)
\end{equation}
for the doublon states. As illustrated in Fig.~(\ref{fig:4+6_hilbert_space}), the last two doublon states have $L=\pm 1$ while the first four doublon states have $L=0$. Now the $\delta U$ term enters as $H'=4\delta U \sum_i L_i^2$.  We also introduce the canonical conjugate $\phi_i$ which has the commutation relation $\left[ \phi_i, L_j \right] = i\delta_{ij}$ with $L_j$. Then in the $\phi$ representation we can write $L$ as $L_i = -i\frac{\partial}{\partial \phi_i}$. The physical electron operators now become

\begin{equation}
    \begin{aligned}
        c_{i;t\sigma} &= \sum_{\sigma^\prime} f^\dagger_{i;b\sigma^\prime}\psi_{i;b\sigma^\prime}\psi_{i;t\sigma} + f^\dagger_{i;t\Bar{\sigma}}\psi_{i;t\Bar{\sigma}}\psi_{i;t\sigma}\varphi_i,\\
        c_{i;b\sigma} &= \sum_{\sigma^\prime}f^\dagger_{i;t\sigma^\prime}\psi_{i;t\sigma^\prime}\psi_{i;b\sigma} + f^\dagger_{i;b\Bar{\sigma}}\psi_{i;b\Bar{\sigma}}\psi_{i;b\sigma} \varphi^\dagger_i,
    \end{aligned}
\end{equation}
where $\varphi_i = \exp\left( -i\phi_i \right)$ is the slave rotor which decreases $L$ by $1$.  Under the internal U(1) gauge transformation associated with $a_\mu$, $\varphi$ doesn't change. Under the internal gauge transformation associated $b_\mu$, $f_i \rightarrow f_i$, $\psi_{i;t} \rightarrow \psi_{i;t}e^{i\theta_b (i)}$, $\psi_{i;b} \rightarrow \psi_{i;b}e^{-i\theta_b (i)}$, $\varphi_i \rightarrow \varphi_i e^{-2i\theta_b (i)}$. So $\varphi$ couples to $-2b_\mu$, which can also be seen from the fact that the time component of $b_\mu$ enforces the constraint $n_{i;\psi_t}-n_{i,\psi_b}-2L_i=0$. Thus $\varphi$ becomes a higgs boson which controls the dynamics of the internal gauge field $b_\mu$. For the global $U(1)$ symmetry, $\varphi$ only change under the gauge transformation associated with $B_\mu$, in which $\psi \rightarrow \psi$, $f_{i;t}\rightarrow f_{i;t}e^{i\theta_d (i)}$, $f_{i;b}\rightarrow f_{i;b}e^{-i\theta_d (i)}$, $\varphi_i \rightarrow \varphi_i e^{2i\theta_d (i)}$. Thereby $\varphi$ couples to $-2b_\mu + 2B_\mu$.

At small $\delta U$, there is no penalty for finite $|L_i|$. So we expect $\varphi$ to condense like the superfluid phase of a boson.  On the other hand, at large $\delta U$, we should have fixed $L_i = 0$, so $\varphi$ should be gapped, like the Mott insulator phase of a boson. Then $
\delta U$ tunes a superfluid to Mott insulator transition of this extra slave rotor $\varphi$.

Our critical theory consists of the  critical theory Eq.~\ref{eq:Lc} tuned by $J_\perp$ together with the superfluid to Mott transition of $\varphi$:

\begin{equation}
    \begin{aligned}
    \mathcal{L}_{\mathrm{tc}} &= \mathcal{L}_{\mathrm{c}} + \left|\left( \partial_\mu - i\left(-2b_\mu+2B_\mu \right) \right) \varphi\right|^2 \\
    &+\Delta_\varphi \left| \varphi \right|^2 + \lambda_3 \left| \varphi \right|^4.
    \end{aligned}
\end{equation}

There is no first order time derivative term on $\varphi$ because under the mirror reflection symmetry which exchange two layers, $\varphi \leftrightarrow \varphi^\dagger$. So the action for $\varphi$ is relativistic, similar to the interaction tuned superfluid to Mott transition of bosons.

 In Fig.~\ref{fig:phase_diagram} we show the schematic zero temperature phase diagram with two tuning parameters: $J_\perp$ which tunes the mass of $\Phi_a$ and $\delta U$ which tunes $\Delta_\varphi$, the gap of $\varphi$. Note that since $\varphi$ couples to $2b$, it contributes to the action of $b_\mu$ and modifies $e_{b,0}^2$ even when $\varphi$ is gapped. In Appendix.~\ref{sec:tri_critical theory} we show that when $\Delta_\varphi > 0$,

\begin{equation}
    \begin{aligned}
        \frac{1}{e_{a,0}^2} &= \frac{1}{6\pi m_\psi} + \frac{2}{3\pi m_f}, \\
        \frac{1}{e_{b,0}^2} &= \frac{1}{6\pi m_\psi} +\frac{1}{6\pi \sqrt{\Delta_\varphi}}.
    \end{aligned}
\end{equation}

At a large $\delta U$, the mass $\Delta_\varphi$ is large, so we still have $e_{a,0}^2 < e_{b,0}^2$ and pairing instability at $\mu_\Phi=0$ as discussed in the previous section. If we decrease $\delta U$ until the gap of $\varphi$ reaches the critical value $\Delta_{\varphi c} = m_f^2 / 16$. Now we have $e_{a,0}^2 = e_{b,0}^2$, which means the repulsive and attractive interactions mediated by gauge fields are balanced at $\mu_\Phi=0$. In this case the pairing instability of the DFFT transition is suppressed. As we further decrease $\delta U$ to get a smaller but still positive $\Delta_\varphi$, we have $e_{a,0}^2 > e_{b,0}^2$. Now when $\mu_\Phi<0$, the $\psi$ Fermi surface is still stable to pairing and we have an intermediate phase with $f, \psi$ Fermi surfaces coupled to deconfined $a_\mu$ and $b_\mu$. This is roughly a stable phase similar to the DQCP and we call it deconfined metal (DM$_1$). Note that in DM$_1$ $f$ couples to $2 a_\mu$ and $\psi$ couples to $a_\mu \pm b_\mu$ as in the DFFT critical regime.  Lastly, if we decrease $\delta U$ until $\Delta_\varphi<0$, we will have $\langle \varphi \rangle \neq 0$ and $b_\mu$ is higgsed completely ( $b_\mu=B_\mu$ ). Then we have a different intermediate deconfined metal (DM$_2$) phase where $f$ couples to $2a_\mu$ and $\psi$ couples to $a_\mu$. The property of the DM phases should be similar to the critical regime of the DFFT discussed in the last section.

\section{conclusion}

In summary, we propose a deconfined quantum critical point (DQCP) between two symmetric Fermi liquids in a bilayer model tuned by inter-layer spin coupling $J_\perp$, with a Fermi surface volume jump of $1/2$ Brillouin zone across the transition. Though the two sides are just conventional Fermi liquids, the critical regime is dominated by fractionalized fermions coupled to two emergent U(1) gauge fields, with the electron operator as a three-particle bound state.   The critical point has an instability towards an intermediate superconductor dome with inter-layer pairing. We also show that tuning another parameter can suppress the pairing instability and lead to a stable deconfined metallic phase in the intermediate regime.  Our phase diagram shows certain similarity to the experimental phase diagram of the hole doped cuprates, with a small to large Fermi surface transition and an associated superconductor dome. But our setup is much cleaner due to the absence of the complexity from various symmetry breaking orders. We hope future investigation of this Fermi liquid to Fermi liquid transition can provide more insights on the general theory of the strange metal and its superconducting instability. The transition may be realized in  the recently found nickelate superconductor La$_3$Ni$_2$O$_7$ tuned by pressure\cite{oh2023type,yang2023strong} and also in quantum simulator based on bilayer optical lattice\cite{bohrdt2021exploration}.

\section{Acknowledgement}

This work was supported by the National Science Foundation under Grant No. DMR-2237031.

\bibliographystyle{apsrev4-1}

\bibliography{ref}

\appendix

\onecolumngrid

\section{Mean field decomposition}
\label{sec:mean_field_decomposition}

In this section we show the mean-field Hamiltonian of Eq.~(\ref{eq:4+4_tJ_model}). Plug the electron operators of parton form Eq.~(\ref{eq:electron_operator}) into Eq.~(\ref{eq:4+4_tJ_model}), we have

\begin{equation}
    \label{eq:parton_hamiltonian}
    \begin{aligned}
        H &= -t\sum_{\langle ij \rangle}\sum_{\sigma,\sigma_1,\sigma_2}\left( \psi^{\dagger}_{i;t\sigma}\psi^{\dagger}_{i;b\sigma_1}f_{i;b\sigma_1}f^{\dagger}_{j;b\sigma_2}\psi_{j;b\sigma_2} \psi_{j;t\sigma}\right. \\
        &+ \left.\psi^{\dagger}_{i;b\sigma}\psi^{\dagger}_{i;t\sigma_1}f_{i;t\sigma_1}f^{\dagger}_{j;t\sigma_2}\psi_{j;t\sigma_2}\psi_{j;b\sigma} \right) \\
        &+ \frac{J_{\perp}}{4}\sum_i \left[ \psi^{\dagger}_{i;t\sigma} \Vec{\sigma}_{\sigma \sigma^{\prime}}\psi_{i;t\sigma^{\prime}} \right]\left[ \psi^{\dagger}_{i;b\sigma}\Vec{\sigma}_{\sigma \sigma^{\prime}}\psi_{i;b\sigma^{\prime}} \right]. 
    \end{aligned}
\end{equation}

We can define the following parameters,
\begin{align}
    \chi_{f;ij}=&\sum_{\sigma}\langle f^\dag_{i;t\sigma}f_{j;t\sigma}\rangle=\sum_{\sigma}\langle f^\dag_{i;b\sigma}f_{j;b\sigma}\rangle,\\
    \chi_{\psi;ij}=&\sum_{\sigma}\langle\psi_{i;t\sigma}^\dag\psi_{j;t\sigma}\rangle=\sum_{\sigma}\langle\psi_{i;b\sigma}^\dag\psi_{j;b\sigma}\rangle,\\
    \chi_{f\psi;i}=&\sum_{\sigma}\langle f_{i;t\sigma}^\dag\psi_{i;t\sigma}\rangle=\sum_{\sigma}\langle f_{i;b\sigma}^\dag\psi_{i;b\sigma}\rangle,\\
    \Tilde{\Delta}_{i}=&\sum_{\sigma\sigma^\prime}\epsilon_{\sigma\sigma^\prime}\langle\psi_{i;t\sigma}\psi_{i;b\sigma^\prime}\rangle=2\langle\psi_{i;t\uparrow}\psi_{b;\downarrow}\rangle=-2\langle\psi_{i;t\downarrow}\psi_{b;\uparrow}\rangle
\end{align}
here for simplicity, we consider all the parameters are in the trivial representation of the lattice symmetry, i.e., $\chi_{ij}=\chi$. In terms of these mean-field parameters, we can use the Wick theorem to obtain the following mean-field Hamiltonian:

\begin{equation}
\begin{aligned}
    H_{\mathrm{MF}}=&\sum_{\langle ij\rangle}\left(t^f f^\dag_{i;a\sigma}f_{j;a\sigma}+\mathrm{h.c.}\right)
    +\sum_{\langle ij \rangle}\left(-t^\psi \psi_{i;l\sigma}^\dag\psi_{j;l\sigma}+\mathrm{h.c.}\right)\\
    +&\sum_{i}\left(-\Phi_a f_{i;a\sigma}^\dag\psi_{i;a\sigma}+\mathrm{h.c.}\right)
    +\sum_{i}\left(-\Delta\epsilon_{\sigma \sigma^\prime}\psi_{i;t\sigma}\psi_{i;b\sigma^\prime} +\mathrm{h.c.}\right)\\
    -&\sum_{i}\left(\mu_{f_t}(n_{i;f_t}-x)+\mu_{f_b}(n_{i;f_b}-x)+\mu_{\psi_t}(n_{\psi_t}-(1-2x))+\mu_{\psi_b}(n_{\psi_b}-(1-2x))\right),
\end{aligned}
\end{equation}

The chemical potentials $\mu_f$ and $\mu_\psi$ are introduced to conserve the particle number. The self consistent equations are:

\begin{equation}
    \begin{aligned}
        t^f &= t\left( \frac{1}{2}\chi_\psi^2 + \frac{1}{4}\Tilde{\Delta}^2 \right), \\
        t^\psi &= t\left( \chi_{f\psi}^2 - \chi_f \chi_\psi \right), \\
        \Phi_a &= t\chi_\psi \chi_{f\psi}, \\
        \Delta &= \left( 2t\chi_f - \frac{3}{8}J_\perp \right)\Tilde{\Delta}.
    \end{aligned}
\end{equation}

\section{Critical theory at large $\delta U$}
\label{sec:critical_theory}

In this section we perform a renormalization group analysis to the critical theory Eq.~(\ref{eq:Lc}), which corresponds to large $\delta U$ case. We can tune $J_\perp$ (the mass $\mu_\Phi$) to reach the QCP at $\mu_\Phi = 0$. We analyze the stability of this QCP while setting the external U(1) gauge fields $A = B = 0$.

\subsection{Self-Energy of the photon}

We basically follow the calculation in \cite{zhang2020pseudogap}. After a renormalization of the gauge field Lagrangian from polarization corrections from the fermion $f$, $\Psi$ and, we obtain:

\begin{equation}
    \mathcal{L}_{a,b} = \frac{1}{2}\left( \frac{1}{e_{a,0}^2}\left| \mathbf{q} \right|^2 + \kappa_a \frac{|\omega|}{|\mathbf{q}|} \right)\left| a(\omega,\mathbf{q}) \right|^2+ \frac{1}{2}\left( \frac{1}{e_{b,0}^2}\left| \mathbf{q} \right|^2 + \kappa_b \frac{|\omega|}{|\mathbf{q}|} \right)\left| b(\omega,\mathbf{q}) \right|^2,
\end{equation}
where we use the Coulomb gauge so there is only one transverse component of each gauge field. The first term comes from the bubble diagrams of $f$ and $\psi$, while the second term only comes from $\psi$. Both terms include a Landau diamagnetic $\mathbf{q}^2$ term and a Landau damping $|\omega|/|\mathbf{q}|$ term. Note that the boson $\Phi_a$ does not contribute to $\mathcal{L}_{a,b}$ since its time derivative is of first order. The coupling constants $e^2$ and the Landau damping coefficients $\kappa$ are

\begin{equation}
    \label{eq:bare_e_square}
    \begin{aligned}
        \frac{1}{e_{a,0}^2} &= \frac{1}{6\pi m_\psi} + \frac{2}{3\pi m_f},\quad \kappa_a = \frac{2m_\psi v_{F,\psi}}{\pi} + \frac{8m_f v_{F,f}}{\pi}, \\
        \frac{1}{e_{b,0}^2} &= \frac{1}{6\pi m_\psi}, \quad\quad\quad~~~\quad\kappa_b = \frac{2m_\psi v_{F,\psi}}{\pi}. \\
    \end{aligned}
\end{equation}

\subsection{RG flow}

We consider the coupling between fermions $f$, $\psi$ and gauge fields $a$,$b$. We use the $\epsilon$-expansion\cite{Metlitski2015} to derive the RG equations. We consider the action

\begin{equation}
    S = S_{a,b}+ S_f + S_{\psi}
\end{equation}
with

\begin{equation}
    \begin{aligned}
        S_{a,b} &= \frac{1}{2}\int \frac{d\omega d^2 q}{(2\pi)^3}\left[ \left( \frac{1}{e_a^2}|q_y|^{1+\epsilon}+\kappa_a \frac{|\omega|}{|q_y|} \right)\left| a(\omega,\mathbf{q}) \right|^2 + \left( \frac{1}{e_b^2}|q_y|^{1+\epsilon}+\kappa_b \frac{|\omega|}{|q_y|} \right) \left| b(\omega,\mathbf{q}) \right|^2 \right], \\
        S_f &= \int \frac{d\omega d^2 k}{(2\pi)^3} \sum_{a=t,b}\Bar{f}_{a;\sigma} \left( i\omega - v_{F,f}k_x - \frac{1}{2m_f}k_y^2 \right)f_{a;\sigma} \\
        &+v_{F,f}\int \frac{d^3 q}{(2\pi)^3}\int \frac{d\omega d^2 k}{(2\pi)^3} \sum_{a=t,b}\Bar{f}_{a;\sigma}(k+q)\cdot 2a(q) f_{a;\sigma}(k), \\
        S_{\psi} &=\int \frac{d\omega d^2 k}{(2\pi)^3}\sum_{a=t,b}\Bar{\psi}_{a;\sigma}\left( i\omega - v_{F,\psi}k_x -\frac{1}{2m_\psi}k_y^2 \right)\psi_{a;\sigma} \\
        &+v_{F,\psi}\int\frac{d^3 q}{(2\pi)^3}\int\frac{d\omega d^2 k}{(2\pi)^3}\Bar{\psi}_{t;\sigma}(k+q)\left( a(q)+b(q) \right)\psi_{t;\sigma}(k) \\
        &+v_{F,\psi}\int\frac{d^3 q}{(2\pi)^3}\int\frac{d\omega d^2 k}{(2\pi)^3}\Bar{\psi}_{b;\sigma}(k+q)\left( a(q)-b(q) \right)\psi_{b;\sigma}(k).
    \end{aligned}
\end{equation}

We have scaling $[k_x]=1$, $[\omega]=1$, $[k_y]=\frac{1}{2}$, $[f]=[\psi] = -\frac{7}{4}$, $[a]=[b] = -\frac{3}{2}$, $[e_a^2] = [e_b^2] = \frac{\epsilon}{2}$, $[m_{f/\psi}] = [v_{F,f/\psi}] = 0$. We define the new coupling constants $\alpha_{a,f} = \frac{e_a^2 v_{F,f}}{4\pi^2}$, $\alpha_{b,f} = \frac{e_b^2 v_{F,f}}{4\pi^2}$, $\alpha_{a,\psi} = \frac{e_a^2 v_{F,\psi}}{4\pi^2}$ and $\alpha_{b,\psi} = \frac{e_b^2 v_{F,\psi}}{4\pi^2}$. The naive scaling gives $[\alpha_a] = [\alpha_b] = \frac{\epsilon}{2}$.

To do the renormalization group analysis using $\epsilon$-expansion, we redefine the fields as follows:

\begin{equation}
    \begin{aligned}
        f^0 &= Z_f^{1/2}f, \\
        \psi^0 &= Z_{\psi}^{1/2} \psi,\\
        v_{F,f}^0 &= Z_{v_{F,f}}v_{F,f}, \\
        v_{F,\psi}^0 &= Z_{v_{F,\psi}}v_{F,\psi},\\
        e_a^0 &= \mu^{\frac{\epsilon}{4}}Z_{e_a}e_a, \\
        e_b^0 &= \mu^{\frac{\epsilon}{4}}Z_{e_b}e_b, \\
        a^0 &= Z_a a, \\
        b^0 &= Z_b b,
    \end{aligned}
\end{equation}
and then we can rewrite the original action as

\begin{equation}
    \begin{aligned}
        S &= \frac{1}{2}\int \frac{d\omega d^2 q}{(2\pi)^3}\left[ \left( \frac{Z_a^2}{\mu^{\frac{\epsilon}{2}}Z_{e_a}^2 e_a^2}|q_y|^{1+\epsilon} + Z_a^2 \kappa_a \frac{|\omega|}{|q_y|} \right)|a(\omega,\mathbf{q})|^2 \right] \\
        &+ \frac{1}{2}\int \frac{d\omega d^2 q}{(2\pi)^3}\left[ \left( \frac{Z_b^2}{\mu^{\frac{\epsilon}{2}}Z_{e_b}^2 e_b^2}|q_y|^{1+\epsilon} + Z_b^2 \kappa_b \frac{|\omega|}{|q_y|} \right)|b(\omega,\mathbf{q})|^2 \right] \\
        &+\int \frac{d\omega d^2 k}{(2\pi)^3}\sum_{a=t,b}\Bar{f}_{a;\sigma}\left( iZ_f \omega - Z_f Z_{v_{F,f}}v_{F,f}k_x -Z_f \frac{1}{2m_f}k_y^2 \right)f_{a;\sigma} \\
        &+Z_f Z_{v_{F,f}}v_{F,f}\int \frac{d^3 q}{(2\pi)^3}\int \frac{d\omega d^2 k}{(2\pi)^3}\sum_{a=t,b}\Bar{f}_{a;\sigma}(k+q)\cdot 2Z_a a(q)f_{a;\sigma}(k) \\
        &+\int \frac{d\omega d^2 k}{(2\pi)^3}\sum_{a=t,b}\Bar{\psi}_{a;\sigma} \left( iZ_\psi\omega - Z_\psi Z_{v_{F,\psi}}v_{F,\psi}k_x -Z_{\psi}\frac{1}{2m_\psi}k_y^2 \right)\psi_{a;\sigma} \\
        &+Z_\psi Z_{v_{F,\psi}}v_{F,\psi}\int \frac{d^3 q}{(2\pi)^3}\int \frac{d\omega d^2 k}{(2\pi)^3}\Bar{\psi}_{t;\sigma}(k+q)\left( Z_a a(q) + Z_b b(q) \right)\psi_{t;\sigma}(k) \\
        &+Z_\psi Z_{v_{F,\psi}}v_{F,\psi}\int \frac{d^3 q}{(2\pi)^3}\int \frac{d\omega d^2 k}{(2\pi)^3}\Bar{\psi}_{b;\sigma}(k+q)\left( Z_a a(q) - Z_b b(q) \right)\psi_{b;\sigma}(k). \\
    \end{aligned}
\end{equation}
From the Ward identity, we expect $Z_a = Z_b = 1$. Hence the fermion-gauge field vertex correction should be purely from $Z_f Z_{v_{F,f}}$ and $Z_\psi Z_{v_{F,\psi}}$. It can be shown that they both equal to 1, implying that there is no vertex correction. When $\epsilon < 1$, we expect $Z_{e_a} = Z_{e_b} = 1$ because the non-analytic form $|q_y|^{1+\epsilon}$ cannot be renormalized. Therefore the only important renormalization is from $Z_f = Z_{v_{F,f}}^{-1}$ and $Z_\psi = Z_{v_{F,\psi}}^{-1}$.

The self-energy of $f$ at one-loop order is

\begin{equation}
    \begin{aligned}
        \Sigma_f (i\omega) &= \frac{4e_a^2 v_{F,f}^2}{(2\pi)^3}\int dq_0 d^2 q \frac{1}{|q_y|^{1+\epsilon} + \kappa_a e_a^2 \frac{|q_0|}{|q_y|}}\frac{1}{i\omega + iq_0 - v_{F,f}(k_x + q_x)- \frac{1}{2m_f}(k_y+q_y)^2} \\
        &=2\alpha_{a,f}\int dq_0 dq_y \frac{i\textrm{sign}(\omega+q_0)}{|q_y|^{1+\epsilon}+\kappa_a e_a^2 \frac{|q_0|}{|q_y|}} \\
        &=4\alpha_{a,f}\frac{1}{\epsilon}\int dq_0 i \textrm{sign}(\omega+q_0) + ... \\
        &=8\alpha_{a,f}i\omega\frac{1}{\epsilon}.
    \end{aligned}
\end{equation}
In the above we only keep the divergent part $\mathcal{O}(1/\epsilon)$. To cancel the divergence, we need

\begin{equation}
    Z_f = Z_{v_{F,f}}^{-1} = 1-8\alpha_{a,f}\frac{1}{\epsilon} + \mathcal{O}(\frac{1}{\epsilon^2}).
\end{equation}

The self-energy of $\psi$ at one-loop order is

\begin{equation}
    \begin{aligned}
        \Sigma_\psi(i\omega) &= \frac{e_a^2 v_{F,\psi}^2}{(2\pi)^3}\int dq_0 d^2 q\frac{1}{|q_y|^{1+\epsilon}+\kappa_a e_a^2 \frac{|q_0|}{|q_y|}}\frac{1}{i\omega + iq_0 - v_{F,\psi}(k_x+q_x)-\frac{1}{2m_\psi}(k_y+q_y)^2} \\
        &+\frac{e_b^2 v_{F,\psi}^2}{(2\pi)^3}\int dq_0 d^2 q\frac{1}{|q_y|^{1+\epsilon}+\kappa_b e_b^2 \frac{|q_0|}{|q_y|}}\frac{1}{i\omega + iq_0 - v_{F,\psi}(k_x+q_x)-\frac{1}{2m_\psi}(k_y+q_y)^2} \\
        &=\frac{\alpha_{a,\psi}}{2}\int dq_0 d^2 q \frac{i\textrm{sign}(\omega+q_0)}{|q_y|^{1+\epsilon}+\kappa_a e_a^2 \frac{|q_0|}{|q_y|}} + \frac{\alpha_{b,\psi}}{2}\int dq_0 d^2 q \frac{i\textrm{sign}(\omega+q_0)}{|q_y|^{1+\epsilon}+\kappa_b e_b^2 \frac{|q_0|}{|q_y|}} \\
        &=\left( \alpha_{a,\psi} + \alpha_{b,\psi} \right)\frac{1}{\epsilon}\int dq_0 i\textrm{sign}(\omega+q_0) + ...\\
        &= 2\left( \alpha_{a,\psi} + \alpha_{b,\psi} \right)i\omega \frac{1}{\epsilon}
    \end{aligned}
\end{equation}
therefore

\begin{equation}
    Z_\psi = Z_{v_{F,\psi}}^{-1} = 1-2\left( \alpha_{a,\psi} + \alpha_{b,\psi} \right)\frac{1}{\epsilon} + \mathcal{O}( \frac{1}{\epsilon^2}).
\end{equation}

Next, we show explicitly that the vertex correction vanishes. For simplicity we use $a\Bar{f}f$ as an illustration. We have

\begin{equation}
    \begin{aligned}
        \delta \Gamma^{c}(p_0, p_x, p_y) &= \int \frac{d^3 q}{2\pi}\frac{1}{iq_0-v_{F,f}q_x - \frac{1}{2K_f}q_y^2}\frac{1}{iq_0 + ip_0 - v_{F,f}(q_x+p_x) - \frac{1}{2K_f}(q_y+p_y)^2} \frac{4\alpha_{a,f}v_{F,f}}{|q_y|^{1+\epsilon}+\kappa_a e_a^2\frac{|q_0|}{|q_y|}} \\
        &=i\textrm{sign}(p_0)\int dq_y \int_0^{|p_0|} dq_0 \frac{4\alpha_{a,f}}{|q_y|^{1+\epsilon}+\kappa_a e_a^2 \frac{|q_0|}{|q_y|}}\frac{1}{ip_0-v_{F,f}p_x - \frac{1}{K_f}p_y q_y - \frac{1}{2K_f}p_y^2},
    \end{aligned}
\end{equation}
where $(p_0, p_x, p_y)$ is the external momentum of $a$ photon at the vertex. We assume one of the external $f$ fermion at the vertex has zero momentum. In the first step we integrate over $q_x$ and get a factor $\textrm{sign}(p_0+q_0) - \textrm{sign}(q_0)$, which equals to 2 for $q_0 \in [-p_0,0]$ and zero otherwise. We can see that $\delta \Gamma^c (p_0 = 0, p_x, p_y) = 0$ so there is no vertex correction. The same conclusion holds for every fermion-gauge field vertex.

Finally, we can get the beta functions $\beta(\alpha) = -d\alpha/d \log \mu$ (note that this is the negative of the usual definition) from the relations

\begin{equation}
    \begin{aligned}
        \alpha_{a,f}^0 &= \mu^{\frac{\epsilon}{2}} Z_{e_a}^2 Z_{v_{F,f}} \alpha_{a,f}, \\
        \alpha_{b,f}^0 &= \mu^{\frac{\epsilon}{2}} Z_{e_b}^2 Z_{v_{F,f}} \alpha_{b,f}, \\
        \alpha_{a,\psi}^0 &= \mu^{\frac{\epsilon}{2}} Z_{e_a}^2 Z_{v_{F,\psi}} \alpha_{a,\psi}, \\
        \alpha_{b,\psi}^0 &= \mu^{\frac{\epsilon}{2}} Z_{e_b}^2 Z_{v_{F,\psi}} \alpha_{b,\psi}. \\
    \end{aligned}
\end{equation}
Using $d\log\alpha^0 /d\log \mu = 0$ and $Z_{e_a} = Z_{e_b} = 1$, we have 

\begin{equation}
    0 = -\frac{\epsilon}{2} - \frac{d\log \alpha}{d\log \mu} - \frac{d\log Z_{v_F}}{d\log\mu} = -\frac{\epsilon}{2} - \frac{d\log \alpha}{d\log \mu} + \frac{d\log Z}{d\log\mu}
\end{equation}
where in the second step we use the relation $Z_f Z_{v_{F,f}} = Z_\psi Z_{v_{F,\psi}} = 1$. We have equations:

\begin{equation}
    \left( 1-\alpha_{a,f}\frac{\partial \log Z_f}{\partial \alpha_{a,f}} \right)\beta (\alpha_{a,f}) = \frac{\epsilon}{2}\alpha_{a,f},
\end{equation}

\begin{equation}
    \beta(\alpha_{b,f}) - \alpha_{b,f}\frac{\partial \log Z_f}{\partial \alpha_{a,f}}\beta(\alpha_{a,f}) = \frac{\epsilon}{2} \alpha_{b,f},
\end{equation}

\begin{equation}
    \left( 1-\alpha_{a,\psi}\frac{\partial \log Z_\psi}{\partial \alpha_{a,\psi}} \right)\beta (\alpha_{a,\psi}) - \alpha_{a,\psi}\frac{\partial \log Z_\psi}{\partial \alpha_{b,\psi}} \beta(\alpha_{b,\psi}) = \frac{\epsilon}{2}\alpha_{a,\psi},
\end{equation}

\begin{equation}
    -\alpha_{b,\psi} \frac{\partial \log Z_\psi}{\partial \alpha_{a,\psi}} \beta(\alpha_{a,\psi}) + \left( 1-\alpha_{b,\psi}\frac{\partial \log Z_\psi}{\partial \alpha_{b,\psi}} \right)\beta(\alpha_{b,\psi}) = \frac{\epsilon}{2}\alpha_{b,\psi}.
\end{equation}
Using

\begin{equation}
    \begin{aligned}
        \log Z_f &= -8\alpha_{a,f}\frac{1}{\epsilon} + \mathcal{O}(\frac{1}{\epsilon^2}), \\
        \log Z_{\psi} &= -2\left( \alpha_{a,\psi} + \alpha_{b,\psi} \right)\frac{1}{\epsilon}+\mathcal{O}(\frac{1}{\epsilon^2}),
    \end{aligned}
\end{equation}
and the fact that all $\mathcal{O}(1/\epsilon^n)$ terms should vanish for the theory to be renormalizable, we obtain the beta functions:

\begin{equation}
    \label{eq:beta function}
    \begin{aligned}
        \beta(\alpha_{a,f}) &= \frac{\epsilon}{2}\alpha_{a,f} - 4\alpha_{a,f}^2 \\
        \beta(\alpha_{b,f}) &= \frac{\epsilon}{2}\alpha_{b,f} - 4\alpha_{a,f}\alpha_{b,f} \\
        \beta(\alpha_{a,\psi}) &= \frac{\epsilon}{2}\alpha_{a,\psi} - \alpha_{a,\psi}\left( \alpha_{a,\psi}+\alpha_{b,\psi}  \right) \\
        \beta(\alpha_{b,\psi}) &= \frac{\epsilon}{2}\alpha_{b,\psi} - \alpha_{b,\psi} \left( \alpha_{a,\psi}+\alpha_{b,\psi} \right)
    \end{aligned}
\end{equation}
We can see that there are fixed points satisfying $\alpha_{a,f} = \epsilon/8$ and $\alpha_{a,\psi} + \alpha_{b,\psi} = \epsilon/2$. We also find that the ratio $\alpha_{a,\psi} / \alpha_{b,\psi}$ doesn't flow, which shows that $\alpha_{a,\psi} - \alpha_{b,\psi}$ doesn't change sign.

\subsection{Pairing instability}

The leading contribution to the interaction in BCS channel for fermions is from exchange of one photon. Among all the fermion pairing terms, the following one is the most important:

\begin{equation}
    S_{BCS} = \int \prod d^2 k_i d\omega_i\Bar{\psi}_t (k_1) \Bar{\psi}_b (-k_1) \psi_b (-k_2) \psi_t (k_2) V F(k_1-k_2),
\end{equation}
where $F(q=k_1-k_2)$ arises from the propagator of photons. Note that $\psi_t$ couples to $a+b$ while $\psi_b$ couples to $a-b$, which means that $a$ mediates repulsive interaction between $\psi_t$ and $\psi_b$ while $b$ mediates attractive interaction. The final sign of the interaction between $\psi_t$ and $\psi_b$ depends on the competition between $a$ and $b$.

We define dimensionless BCS interaction constant

\begin{equation}
    \Tilde{V}_{m} = \frac{k_{F,\psi}}{2\pi v_{F,\psi}}V_m,
\end{equation}
where $m$ is the angular momentum for the corresponding pairing channel. By integrating out photon in the intermediate energy, we obtain

\begin{equation}
    \label{eq:rg_V}
    \begin{aligned}
        \delta \Tilde{V}_m &= \frac{k_{F,\psi}}{2\pi v_{F,\psi}}v_{F,\psi}^2 \int \frac{d\theta}{2\pi} \left( \frac{e^{-im\theta}}{|k_{F,\psi}\theta|^{1+\epsilon}/e_a^2} - \frac{e^{-im\theta}}{|k_{F,\psi}\theta|^{1+\epsilon}/e_b^2} \right) \\
        &=\frac{v_{F,\psi}}{4\pi^2}2\int_{\Lambda_y e^{-\delta l/2}}^{\Lambda_y} dq_y \left( \frac{1}{|q_y|^{1+\epsilon}/e_a^2} - \frac{1}{|q_y|^{1+\epsilon}/e_b^2} \right) \\
        &=\left( \alpha_{a,\psi} - \alpha_{b,\psi} \right)\delta l.
    \end{aligned}
\end{equation}

The renormalization in Eq.~(\ref{eq:rg_V}) should be combined with the usual flow of the BCS interaction to obtain the RG equation:

\begin{equation}
    \label{eq:beta function_V}
    \frac{d\Tilde{V}}{dl} = \left( \alpha_{a,\psi} - \alpha_{b,\psi} \right) - \Tilde{V}^2.
\end{equation}
From Eq.~(\ref{eq:bare_e_square}) we have bare values $\alpha^0_{a,\psi} - \alpha^0_{b,\psi} < 0$, then $\Tilde{V}$ will flow to $-\infty$. Considering the case $\epsilon = 0$, then from Eq.~(\ref{eq:beta function}) we obtain

\begin{equation}
    \begin{aligned}
        \alpha_{a,\psi}(l) &= \frac{\alpha_{a,\psi}(0)}{1+\left( \alpha_{a,\psi}(0)+\alpha_{b,\psi}(0) \right)l}, \\
        \alpha_{b,\psi}(l) &= \frac{\alpha_{b,\psi}(0)}{1+\left( \alpha_{a,\psi}(0)+\alpha_{b,\psi}(0) \right)l},
    \end{aligned}
\end{equation}
where we identify $-\log \mu$ as $l$. The decreasing of $\alpha$ is slow so we can approximately use the bare value when solving Eq.~(\ref{eq:beta function_V}). We get

\begin{equation}
    \Tilde{V}(l) = \sqrt{\alpha_{b,\psi}^0-\alpha_{a,\psi}^0}\tan \left( -\sqrt{\alpha_{b,\psi}^0-\alpha_{a,\psi}^0} l +\tan^{-1}\frac{\Tilde{V}(0)}{\sqrt{\alpha_{b,\psi}^0-\alpha_{a,\psi}^0}} \right).
\end{equation}
We can see that $\Tilde{V}(l)$ diverges at 

\begin{equation}
    l_p = \frac{1}{\sqrt{\alpha_{b,\psi}^0 - \alpha_{a,\psi}^0}}\left( \frac{\pi}{2}+\tan^{-1}\frac{\Tilde{V}(0)}{\sqrt{\alpha_{b,\psi}^0-\alpha_{a,\psi}^0}} \right),
\end{equation}
which also gives a estimation of the superconducting gap $\Delta \sim \Lambda_{\omega}\exp(-l_p)$. In the limit $\Tilde{V}(0) \ll \sqrt{\alpha_{b,\psi}^0-\alpha_{a,\psi}^0}$ , we have $l_p \approx \pi / 2\sqrt{\alpha_{b,\psi}^0-\alpha_{a,\psi}^0}$. In the limit $\Tilde{V}(0) \gg \sqrt{\alpha_{b,\psi}^0 - \alpha_{a,\psi}^0}$, we have $l_p \approx \pi / \sqrt{\alpha_{b,\psi}^0-\alpha_{a,\psi}^0}$. Here $\Lambda_\omega$ is the high energy cutoff of RG.

\section{Pairing instability in the critical region}
\label{sec:tri_critical theory}

Now the photon self-energy is given by the polarization corrections from the fermion $f$, $\psi$ and boson $\varphi$. The contribution from $\varphi$ bubble is

\begin{equation}
    \begin{aligned}
        \delta\Pi^{\mu\nu}(q) &= 4\int \frac{d^3 l}{(2\pi)^3}\left[ \frac{(2l+q)^\mu (2l+q)^\nu}{\left( (l+q)^2 +\Delta_\varphi \right)\left( l^2 + \Delta_\varphi \right)} - \frac{2\delta^{\mu\nu}}{l^2 + \Delta_\varphi} \right] \\
        &= \int_{-\frac{1}{2}}^{\frac{1}{2}} dy \frac{2y^2}{\pi\sqrt{\left( \frac{1}{4}-y^2 \right)q^2+\Delta_\varphi}} \left(q^2\delta^{\mu\nu} - q^\mu q^\nu \right).
    \end{aligned}
\end{equation}
The factor 4 comes from the fact that $\varphi$ couples to $-2b_\mu$. The Feynman parametrization is used to get the second line. It has different behaviors for $\Delta_\varphi > 0$ and $\Delta_\varphi = 0$:

\begin{equation}
    \begin{aligned}
        &\Delta_\varphi = 0:~~~ \delta\Pi^{\mu\nu} = \frac{q^2 \delta^{\mu\nu}-q^\mu q^\nu}{4\left| q \right|}, \\
        &\Delta_\varphi > 0:~~~ \delta\Pi^{\mu\nu} =\frac{q^2 \delta^{\mu\nu}-q^{\mu}q^\nu}{6\pi \sqrt{\Delta_\varphi}} + \mathcal{O}(q^4).
    \end{aligned}
\end{equation}
Combining all the polarization corrections from $f$, $\psi$ and $\varphi$, for $\Delta_\varphi > 0$ and $\mu_\Phi = 0$ we have

\begin{equation}
    \mathcal{L}_{a,b} = \frac{1}{2}\left( \frac{1}{e_{a,0}^2}\left| \mathbf{q} \right|^2 + \kappa_a \frac{\left| \omega \right|}{\left| \mathbf{q} \right|} \right) \left| a(\omega,\mathbf{q}) \right|^2 + \frac{1}{2}\left( \frac{1}{e_{a,0}^2}\left| \mathbf{q} \right|^2 + \kappa_a \frac{\left| \omega \right|}{\left| \mathbf{q} \right|} \right) \left| b(\omega,\mathbf{q}) \right|^2,
\end{equation}
where the coupling constants $e^2$ are 

\begin{equation}
    \label{eq:bare_e_square_deltaU}
    \begin{aligned}
        \frac{1}{e_{a,0}^2} &= \frac{1}{6\pi m_\psi} + \frac{2}{3\pi m_f}, \\
        \frac{1}{e_{b,0}^2} &= \frac{1}{6\pi m_\psi} + \frac{1}{6\pi\sqrt{ \Delta_\varphi}}.
    \end{aligned}
\end{equation}
Note that all the RG analysis in Appendix.~\ref{sec:critical_theory} is still valid when $\varphi$ is gapped. At $\Delta_\varphi = \Delta_{\varphi c} = \frac{1}{16}m_f^2$, we have $e_{b,0}^2 = e_{a,0}^2$. The attractive and repulsive interactions mediated by gauge fields are balanced.

\section{calculation of Green's function}

Consider the electron Green's function $G_c(\mathbf{k},i\omega) = -\langle c_{t\uparrow}(\mathbf{k}, i\omega) c^\dagger_{t\uparrow}(\mathbf{k}, i\omega) \rangle$. In the following, I use $\xi_f(\mathbf{k}) = \epsilon_f (\mathbf{k})-\mu_f$, $\xi_\psi (\mathbf{k}) = \epsilon_\psi (\mathbf{k})-\mu_\psi$, $k = (\mathbf{k}, i\omega)$. From the parton construction Eq.~\ref{eq:electron_operator}, we have

\begin{equation}
    \begin{aligned}
        G_c(k) &= -2\int_{k_1}\int_{k_2}\langle f^\dagger_{b\uparrow}(-k_2) f_{b\uparrow}(-k_2) \rangle \langle \psi_{b\uparrow}(k_1) \psi^\dagger_{b\uparrow} (k_1) \rangle \langle \psi_{t\uparrow}(k-k_1-k_2)\psi^\dagger_{t\uparrow}(k-k_1-k_2) \rangle \\
        &=2\int_{k_1}\int_{k_2}\frac{-1}{-i\omega_2 - \xi_f(\mathbf{k}_2)}\cdot \frac{1}{i\omega_1 - \xi_\psi(\mathbf{k}_1)}\cdot \frac{1}{i\omega - i\omega_1 - i\omega_2 - \xi_\psi (\mathbf{k} - \mathbf{k}_1 - \mathbf{k}_2)} \\
        &= 2\int_{k_1}\int_{\mathbf{k}_2} \frac{1}{i\omega_1 - \xi_\psi(\mathbf{k}_1)} \cdot \frac{\Theta(\xi_\psi(\mathbf{k}-\mathbf{k}_1 - \mathbf{k}_2)) - \Theta(\xi_f (\mathbf{k}_2))}{-i\omega_1 + i\omega + \xi_f(\mathbf{k}_2) - \xi_\psi(\mathbf{k}-\mathbf{k}_1 - \mathbf{k}_2)} \\
        &=2\int_{\mathbf{k}_1}\int_{\mathbf{k}_2}\frac{\left( \Theta(\xi_f(\mathbf{k}_2) - \xi_\psi(\mathbf{k}-\mathbf{k}_1 - \mathbf{k}_2)) - \Theta(\xi_\psi(\mathbf{k}_1)) \right) \left( \Theta(\xi_f(\mathbf{k}_2)) - \Theta(\xi_\psi(\mathbf{k}-\mathbf{k}_1 - \mathbf{k}_2)) \right)}{i\omega + \xi_f(\mathbf{k}_2) - \xi_\psi(\mathbf{k}-\mathbf{k}_1 - \mathbf{k}_2) - \xi_\psi(\mathbf{k}_1)} \\
        &=2\int_{\mathbf{k}_1}\int_{\mathbf{k}_2}\frac{\left[ \Theta(\xi_f(\mathbf{k}_2))\Theta(-\xi_\psi(\mathbf{k}_1))\Theta(-\xi_\psi(\mathbf{k}-\mathbf{k}_1-\mathbf{k}_2)) +  \Theta(-\xi_f(\mathbf{k}_2))\Theta(\xi_\psi(\mathbf{k}_1))\Theta(\xi_\psi(\mathbf{k}-\mathbf{k}_1-\mathbf{k}_2))\right]}{i\omega + \xi_f(\mathbf{k}_2) - \xi_\psi(\mathbf{k}-\mathbf{k}_1 - \mathbf{k}_2)-\xi_\psi(\mathbf{k}_1)}.
    \end{aligned}
\end{equation}

The spectral function is then

\begin{equation}
    A(\omega, \mathbf{k}) = -\frac{1}{\pi}\text{Im} G^R_c(\omega, \mathbf{k}) = -\frac{1}{\pi}\text{Im} G_c(k)\large|_{i\omega \rightarrow \omega + i\eta}.
\end{equation}

\subsection{Green's function in sFL}

The mean-field Hamiltonian of sFL is

\begin{equation}
    H = \sum_{\mathbf{k}} \xi_f(\mathbf{k})f^\dagger_{a\sigma}(\mathbf{k})f_{a\sigma}(\mathbf{k}) + \sum_{\mathbf{k} \alpha}\Psi^{(\alpha)\dagger}(\mathbf{k})h(\mathbf{k})\Psi^{(\alpha)}(\mathbf{k}),
\end{equation}
where $\alpha = 1,2$, 

\begin{equation}
    \begin{aligned}
        \Psi^{(1)} = \left( \psi_{t\uparrow}(\mathbf{k}), \psi_{b\downarrow}^\dagger (-\mathbf{k}) \right)^T, \\
        \Psi^{(2)} = \left( \psi_{b\uparrow}(\mathbf{k}), \psi_{t\downarrow}^\dagger (-\mathbf{k}) \right)^T,
    \end{aligned}
\end{equation}
and

\begin{equation}
    h(\mathbf{k}) = \begin{pmatrix}
        \xi_\psi(\mathbf{k}) & \Delta \\
        \Delta & -\xi_\psi(\mathbf{k})
    \end{pmatrix}.
\end{equation}

Now the electron Green's function becomes

\begin{equation}
    \begin{aligned}
        G_c(k) &= -2 \int_{k_1}\int_{k_2}\langle f^\dagger_{b\uparrow}(-k_2) f_{b\uparrow}(-k_2) \rangle \langle \psi_{b\uparrow}(k_1) \psi^\dagger_{b\uparrow}(k_1) \rangle \langle \psi_{t\uparrow}(k-k_1-k_2) \psi^\dagger_{t\uparrow}(k-k_1-k_2) \rangle \\
        &-\langle f_{b\downarrow}^\dagger(-k)f_{b\downarrow}(-k)  \rangle \int_{k_1} \langle \psi_{b\downarrow}(k_1)\psi_{t\uparrow}(-k_1) \rangle \int_{k_2} \langle \psi_{t\uparrow}^\dagger(-k_2) \psi_{b\downarrow}^\dagger(k_2) \rangle \\
        &= 2\int_{k_1}\int_{k_2}\frac{-1}{-i\omega_2 - \xi_f(\mathbf{k}_2)} \cdot \frac{i\omega_1 + \xi_\psi(\mathbf{k}_1)}{(i\omega_1)^2 - E^2_\psi(\mathbf{k}_1)}\frac{i\omega - i\omega_1 - i\omega_2 + \xi_\psi(\mathbf{k}-\mathbf{k}_1 - \mathbf{k}_2)}{(i\omega - i\omega_1 - i\omega_2)^2 - E^2_\psi(\mathbf{k} - \mathbf{k}_1 - \mathbf{k}_2)} \\
        &+ \frac{1}{i\omega + \xi_f(\mathbf{k})}\left( \int_{k_1} \frac{\Delta}{(i\omega_1)^2 - E^2_\psi(\mathbf{k}_1)} \right)^2
    \end{aligned}
\end{equation}
where $\xi(\mathbf{k}) = \epsilon(\mathbf{k})-\mu$, $E_\psi(\mathbf{k}) = +\sqrt{\xi_\psi^2(\mathbf{k})+\Delta^2}$. To simplify calculation, we can use

\begin{equation}
    \begin{aligned}
    &-\langle \psi_{b\uparrow}(k) \psi^\dagger_{b\uparrow}(k) \rangle = \frac{i\omega + \xi_\psi(\mathbf{k})}{(i\omega)^2 - E_\psi^2(\mathbf{k})}=\frac{u^2(\mathbf{k})}{i\omega - E_\psi(\mathbf{k})} + \frac{v^2(\mathbf{k})}{i\omega + E_\psi(\mathbf{k})}\\
    &-\langle \psi_{b\uparrow}(k) \psi_{t\downarrow}(-k) \rangle = \frac{\Delta}{(i\omega)^2 - E_\psi^2 (\mathbf{k})} = \frac{\Delta}{2E_\psi(\mathbf{k})}\left( \frac{1}{i\omega - E_\psi(\mathbf{k})} - \frac{1}{i\omega + E_\psi(\mathbf{k})} \right)
    \end{aligned}
\end{equation}
where

\begin{equation}
    u^2(\mathbf{k}) = \frac{1}{2}\left( 1+\frac{\xi_\psi(\mathbf{k})}{E_\psi(\mathbf{k})} \right), ~~~v^2(\mathbf{k}) = \frac{1}{2}\left( 1-\frac{\xi_\psi(\mathbf{k})}{E_\psi(\mathbf{k})} \right).
\end{equation}
Then

\begin{equation}
    \begin{aligned}
        G_c(k) &= 2\int_{k_1} \int_{k_2}\frac{-1}{-i\omega_2 - \xi_f(\mathbf{k}_2)}\frac{1}{i\omega_1 - E_\psi(\mathbf{k}_1)}\frac{u^2(\mathbf{k}_1)u^2(\mathbf{k}-\mathbf{k}_1-\mathbf{k}_2)}{i\omega - i\omega_1 - i\omega_2 - E_\psi(\mathbf{k}-\mathbf{k}_1 - \mathbf{k}_2)} \\
        &~~~~~~~~~~~~ + \frac{-1}{-i\omega_2-\xi_f(\mathbf{k}_2)}\frac{1}{i\omega_1 - E_\psi(\mathbf{k}_1)}\frac{u^2(\mathbf{k}_1)v^2(\mathbf{k}-\mathbf{k}_1-\mathbf{k}_2)}{i\omega - i\omega_1 - i\omega_2 + E_\psi(\mathbf{k}-\mathbf{k}_1 - \mathbf{k}_2)} \\
        &~~~~~~~~~~~~ + \frac{-1}{-i\omega_2-\xi_f(\mathbf{k}_2)}\frac{1}{i\omega_1 + E_\psi(\mathbf{k}_1)}\frac{v^2(\mathbf{k}_1)u^2(\mathbf{k}-\mathbf{k}_1-\mathbf{k}_2)}{i\omega - i\omega_1 - i\omega_2 - E_\psi(\mathbf{k}-\mathbf{k}_1 - \mathbf{k}_2)} \\
        &~~~~~~~~~~~~ + \frac{-1}{-i\omega_2-\xi_f(\mathbf{k}_2)}\frac{1}{i\omega_1 + E_\psi(\mathbf{k}_1)}\frac{v^2(\mathbf{k}_1)v^2(\mathbf{k}-\mathbf{k}_1-\mathbf{k}_2)}{i\omega - i\omega_1 - i\omega_2 + E_\psi(\mathbf{k}-\mathbf{k}_1 - \mathbf{k}_2)} \\
        & + \frac{1}{i\omega + \xi_f(\mathbf{k})}\left( \int_{k_1} \frac{\Delta}{2E_\psi(\mathbf{k}_1)}\left( \frac{1}{i\omega_1 - E_\psi(\mathbf{k}_1)} - \frac{1}{i\omega_1 + E_\psi(\mathbf{k}_1)} \right) \right)^2 \\
        & (\mathbf{k_3} = \mathbf{k} - \mathbf{k}_1 - \mathbf{k}_2) \\
        &=2\int_{\mathbf{k}_1}\int_{\mathbf{k}_2}\left( \frac{u^2(\mathbf{k}_1) u^2(\mathbf{k}_3)\Theta(-\xi_f(\mathbf{k}_2))}{i\omega + \xi_f(\mathbf{k}_2) - E_\psi(\mathbf{k}_1) - E_\psi(\mathbf{k}_3)} + \frac{v^2(\mathbf{k}_1)v^2(\mathbf{k}_3)\Theta(+\xi_f(\mathbf{k}_2))}{i\omega + \xi_f(\mathbf{k}_2) + E_\psi(\mathbf{k}_1) + E_\psi(\mathbf{k}_3)} \right) \\
        &+ \frac{1}{i\omega + \xi_f(\mathbf{k})}\left( \int_{\mathbf{k}_1} \frac{\Delta}{2E_\psi(\mathbf{k}_1)} \right)^2
    \end{aligned}
\end{equation}

\subsection{Green's function in FL}
The mean-field Hamiltonian of FL is

\begin{equation}
    \begin{aligned}
    H &= \sum_{\mathbf{k}}\xi_f (\mathbf{k})f^\dagger_{a\sigma}(\mathbf{k})f_{a\sigma}(\mathbf{k}) + \xi_\psi(\mathbf{k})\psi^\dagger_{a\sigma}(\mathbf{k})\psi_{a\sigma}(\mathbf{k})- \Phi f^\dagger_{a\sigma}(\mathbf{k})\psi_{a\sigma}(\mathbf{k}) - \Phi\psi^\dagger_{a\sigma}(\mathbf{k})f_{a\sigma}(\mathbf{k}) \\
    &= \begin{pmatrix}
        f^\dagger_{a\sigma}(\mathbf{k}), & \psi^\dagger_{a\sigma}(\mathbf{k})
    \end{pmatrix}h(k)
    \begin{pmatrix}
        f_{a\sigma}(\mathbf{k}) \\
        \psi_{a\sigma}(\mathbf{k})
    \end{pmatrix},
    \end{aligned}
\end{equation}
with

\begin{equation}
    h(\mathbf{k}) = \begin{pmatrix}
        \xi_f(\mathbf{k}) & -\Phi \\
        -\Phi & \xi_\psi(\mathbf{k})
    \end{pmatrix}.
\end{equation}
We have

\begin{equation}
    -\begin{pmatrix}
        \langle f_{a\sigma}(k) f^\dagger_{a\sigma}(k) \rangle & \langle f_{a\sigma}(k) \psi^\dagger_{a\sigma}(k) \rangle \\
        \langle \psi_{a\sigma}(k) f^\dagger_{a\sigma}(k) \rangle & \langle \psi_{a\sigma}(k) \psi^\dagger_{a\sigma}(k) \rangle
    \end{pmatrix}
    =\frac{1}{(i\omega - E_{+}(\mathbf{k}))(i\omega - E_{-}(\mathbf{k}))} 
    \begin{pmatrix}
        i\omega - \xi_\psi(\mathbf{k}) & -\Phi \\
        -\Phi & i\omega - \xi_f(\mathbf{k})
    \end{pmatrix},
\end{equation}
where 

\begin{equation}
    E_{\pm} = \frac{\xi_f + \xi_\psi}{2} \pm \sqrt{\left( \frac{\xi_f - \xi_\psi}{2} \right)^2 + \Phi^2}.
\end{equation}
The following equations are useful:

\begin{equation}
    \begin{aligned}
        -\langle f_{a\sigma}(k) f^\dagger_{a\sigma}(k) \rangle &= \frac{\xi_f (\mathbf{k}) - E_{-}(\mathbf{k})}{E_{+}(\mathbf{k}) - E_{-}(\mathbf{k})}\frac{1}{i\omega - E_{-}(\mathbf{k})} + \frac{E_{+}(\mathbf{k}) - \xi_f(\mathbf{k})}{E_{+}(\mathbf{k})- E_{-}(\mathbf{k})}\frac{1}{i\omega - E_{+}(\mathbf{k})} = \frac{u^2_f(\mathbf{k})}{i\omega - E_{+}(\mathbf{k})} + \frac{1-u^2_f(\mathbf{k})}{i\omega - E_{-}(\mathbf{k})}, \\
        -\langle \psi_{a\sigma}(k) \psi^\dagger_{a\sigma}(k) \rangle &= \frac{\xi_\psi (\mathbf{k}) - E_{-}(\mathbf{k})}{E_{+}(\mathbf{k}) - E_{-}(\mathbf{k})}\frac{1}{i\omega - E_{-}(\mathbf{k})} + \frac{E_{+}(\mathbf{k}) - \xi_\psi(\mathbf{k})}{E_{+}(\mathbf{k})- E_{-}(\mathbf{k})}\frac{1}{i\omega - E_{+}(\mathbf{k})} = \frac{u^2_\psi (\mathbf{k})}{i\omega - E_{+}(\mathbf{k})} + \frac{1-u^2_\psi(\mathbf{k})}{i\omega - E_{-}(\mathbf{k})}, \\
        -\langle f_{a\sigma}(k) \psi^\dagger_{a\sigma}(k) \rangle &= \frac{-\Phi}{E_{+}(\mathbf{k}) - E_{-}(\mathbf{k})}\left( \frac{1}{i\omega - E_{+}(\mathbf{k})} - \frac{1}{i\omega - E_{-}(\mathbf{k})} \right) = u_{f\psi}(\mathbf{k})\left( \frac{1}{i\omega - E_{+}(\mathbf{k})} - \frac{1}{i\omega - E_{-}(\mathbf{k})} \right), \\
        -\langle \psi_{a\sigma}(k) f^\dagger_{a\sigma}(k) \rangle &= \frac{-\Phi}{E_{+}(\mathbf{k}) - E_{-}(\mathbf{k})}\left( \frac{1}{i\omega - E_{+}(\mathbf{k})} - \frac{1}{i\omega - E_{-}(\mathbf{k})} \right) = u_{f\psi}(\mathbf{k})\left( \frac{1}{i\omega - E_{+}(\mathbf{k})} - \frac{1}{i\omega - E_{-}(\mathbf{k})} \right).
    \end{aligned}
\end{equation}

Now the electron Green's function becomes

\begin{equation}
    \begin{aligned}
        G_c(k) &= -2\int_{k_1}\int_{k_2}\langle f^\dagger_{b\uparrow}(-k_2) f_{b\uparrow}(-k_2) \rangle \langle \psi_{b\uparrow}(k_1) \psi^\dagger_{b\uparrow}(k_1) \rangle \langle \psi_{t\uparrow}(k-k_1-k_2) \psi^\dagger_{t\uparrow}(k-k_1-k_2) \rangle \\
        &~~~ -4\langle \psi_{t\uparrow}(k) \psi^\dagger_{t\uparrow}(k) \rangle\int_{k_1}\langle f^\dagger_{b\uparrow}(k_1) \psi_{b\uparrow}(k_1) \rangle\int_{k_2} \langle \psi^\dagger_{b\uparrow}(k_2)f_{b\uparrow}(k_2) \rangle \\
        &= 2\int_{\mathbf{k}_1}\int_{\mathbf{k}_2}u_f^2(\mathbf{k}_2)u^2_\psi(\mathbf{k}_1)u^2_\psi(\mathbf{k}_3)\frac{\Theta(E_{+}(\mathbf{k}_2))\Theta(-E_{+}(\mathbf{k}_1))\Theta(-E_{+}(\mathbf{k}_3))+\Theta(-E_{+}(\mathbf{k}_2))\Theta(E_{+}(\mathbf{k}_1))\Theta(E_{+}(\mathbf{k}_3))}{i\omega+E_{+}(\mathbf{k}_2) - E_{+}(\mathbf{k}_1) - E_{+}(\mathbf{k}_3)} \\
        &~~~~~~~~~~~~ + \left( \text{other 7 terms with } u^2 \rightarrow 1-u^2, E_{+} \rightarrow E_{-} \right) \\
        &+4\left( \frac{u^2_\psi(\mathbf{k})}{i\omega - E_{+}(\mathbf{k})} + \frac{1-u^2_\psi(\mathbf{k})}{i\omega - E_{-}(\mathbf{k})} \right) \left( \int_{\mathbf{k}_1}u_{f\psi}(\mathbf{k}_1)\Theta(\Phi^2 - \xi_f(\mathbf{k}_1)\xi_\psi(\mathbf{k}_1)) \right)^2
    \end{aligned}
\end{equation}

\end{document}